\documentclass[final,5p,times,twocolumn]{elsarticle}

\usepackage{amssymb}
\usepackage{lipsum}
\usepackage{amsmath} 
\usepackage{bbm}
\usepackage[normalem]{ulem}

\usepackage{xcolor}

\usepackage{hyperref}
\usepackage{doi}

\usepackage[switch]{lineno}

\journal{High Energy Astrophysics}

\begin{document}

\begin{frontmatter}



\title{Slingshot on compact binaries in the nuclear region of galaxies as the origin of large offset gamma-ray bursts}


\author[first]{Juan C. Rodr\'iguez-Ram\'irez}
            
\author[first]{Viviane Alfradique}

\author[first]{Clecio R. Bom}

\author[second]{Davi C. Rodrigues}

\affiliation[first]{organization={Centro Brasileiro de Pesquisas F\'isicas},
            addressline={Xavier Sigaud st. 150}, 
            city={Rio de Janeiro},
            postcode={22290-180}, 
            state={RJ},
            country={Brazil}}

\affiliation[second]{organization={Departamento de Física \& Núcleo Cosmo-Ufes, Universidade Federal do Espírito Santo, 29075-910 Vitória–ES},
country={Brazil}     }


\begin{abstract}
A growing number of gamma-ray bursts (GRBs) have been observed at large projected offsets from the centres of their putative host galaxies, sometimes extending to several tens of kiloparsecs.
When the GRB is associated with a kilonova, a compact binary with at least one neutron star is expected to be involved, and the cause of the offset can be attributed to the neutron star natal kick.
In this work, we explore an alternative mechanism: the gravitational scattering of compact binaries 
populating the nuclear region of the host galaxy,
by an intermediate-mass black hole or secondary supermassive black hole.
We specifically consider populations of neutron star--neutron star (NS-NS), neutron star--black hole (NS-BH), and neutron star--white dwarf (NS-WD) binaries, and analyse the ejection of bound systems from the nuclear region through a gravitational-slingshot effect.
We proceed analytically to derive the conditions for binary survival and to parametrise the post-encounter velocity.
To quantify the associated merger offsets, we numerically integrate the post-slingshot trajectory of the binary centre-of-mass in host-galaxy potentials that include the primary SMBH, the stellar component, and the dark-matter halo, considering both elliptical-like and spiral-like galaxy profiles.
Our parameter sampling shows that
NS--NS and NS--WD binaries typically merge farther from the
galactic centre than NS--BH systems, with approximately 30--45\% of successful ejections
merging at three-dimensional distances exceeding 100 kpc; these extreme-offset mergers occur within $\sim$0.1--10 Gyr after ejection.
These results motivate the analysed
mechanism as a candidate channel for observed extreme-offset transients.
\end{abstract}




\end{frontmatter}



\section{Introduction}
\label{introduction}

Binary systems of compact objects involving neutron stars,
such as neutron star--neutron star (NS--NS), and neutron star--black hole (NS--BH) binaries, 
have emerged as key sources in modern astrophysics and cosmology.
The coalescence of such binaries, is expected to produce multiwavelength and multimessenger transients, including gamma-ray bursts (GRBs) \cite{Goldstein2017}, gravitational waves (GWs)\cite{Abbott2017}, kilonovae \cite{Metzger2010}, and possibly high-energy neutrinos \cite{Bartos_2013, Kimura_2017}.
These signals provide unique opportunities to probe fundamental physics, from matter under extreme conditions \citep{Janiuk2014, Kasen2017, Coughlin2019, Dietrich2020, Chen2024} to independent measurements of cosmological parameters \citep{schutz, Holz2005, Dalal2006, 2017Natur.551...85A}.

Recent observations of peculiar GRBs are providing valuable, but still not fully understood, information about the compact-binary progenitor population.
Examples of GRBs associated with kilonova signatures, such as GRB 211211A \citep{Rastinejad_2022} and GRB 230307A \citep{Yang_2024Natur,Levan_2024Natur}, suggest that compact binary coalescences can produce long duration GRBs.
This challenges the phenomenological dichotomy in which long GRBs, with durations $\gtrsim 2$ s, are traditionally associated with the collapse of massive stars, whereas short GRBs, lasting $\lesssim 2$ s, are attributed to compact binary mergers \citep{Berger2014}.
At the same time, a growing number of GRBs, both short and long, have been observed at large projected offsets from the centres of their putative host galaxies
\citep{Bloom_2006,Cenko_2008,Fong_2013,Fong2022,Levan_2024Natur,Yang_2024Natur}.
In the most extreme cases, the offsets exceed several tens of kiloparsecs
\citep{Bloom_2006,Bloom2007,Cenko_2008,Levan_2024Natur},
challenging the identification of the transient host galaxy \citep[e.g.][]{Troja2026}.

A commonly invoked mechanism for explaining offset GRBs is 
a compact binary progenitor in which 
one of the components undergo a natal kick during an asymmetric supernova explosion \citep{Kalogera_1996}. Depending on the magnitude and direction, such kick can impart a systemic velocity of tens to a few hundred km~s$^{-1}$ to the surviving binary \citep{Wongwathanarat_2013}, displacing it from its birth site before merger. However, stronger kicks also increase the probability of binary disruption \citep{Tauris_1998,Tauris_2017}, thereby limiting the offsets that compact bound binaries can achieve through this channel.

This limitation has recently been highlighted by GRB~230307A, a long GRB with a claimed kilonova association, and with a projected offset of $\sim39$ kpc from the centre of the closest galaxy \cite{Levan_2024Natur,Yang_2024Natur}.
Detailed analysis of such large offset transient indicate that, if originated in a merger event, a natal kick interpretation is possible only for fine-tuned binary configurations \citep{Alfradique2025, Bom2026}, challenging our understanding of the birth environment of the progenitor.

Motivated by this peculiar event and by the limitations of natal kicks in producing large offset GRBs, in this work we explore an alternative mechanism for the ejection of compact binaries.
We consider the scenario of compact binaries with at least one neutron star 
(hereafter ``NSX binaries'') that populate the nuclear region of the host galaxy.
Then, we examine the systemic velocities that such binaries acquire through gravitational slingshots, when encountering a much more massive body.
We specifically consider the scatterer to be an intermediate-mass black hole (IMBH) or a secondary SMBH, and focus on NSX systems such as NS-NS, NS-BH, and neutron star-white dwarf (NS-WD) binaries.
In this situation, the centre-of-mass (COM) of a NSX binary can be accelerated above the local escape speed from the central SMBH while remaining internally bound.
Since the escape speed in the influence sphere of central SMBHs can range from hundreds to thousands of km s$^{-1}$, an NSX binary launched with such velocities can travel far from the nuclear region and, depending on its merger time, potentially coalesce beyond the main observed stellar body of the host galaxy.

Several non-exclusive channels can seed the central pc of galaxies with compact remnants and hard compact binaries. In low- and intermediate-mass galaxies, nuclear star clusters (NSCs) provide dense stellar environments in which compact objects can be retained and dynamically processed through exchange interactions \citep{FaucherGiguere2011,Sedda2020} and tidal captures \citep{Generozov2018}.
NSCs themselves are thought to form through a combination of in-situ star formation and the infall of globular clusters (GCs) from larger galactocentric radii \citep{Neumayer2020}. Since GCs can form compact binaries involving neutron stars \citep{Ivanova2008,Benacquista2013}, their infall provides a mechanism for transporting compact remnants and hard binaries into the nuclear region. In more massive galaxies, where present day NSCs may be absent, the legacy of previous galaxy mergers can contribute to the compact-object reservoir at sub-pc scales, since galaxy mergers can produce triaxial remnants which in turn sustain a supply of stars and compact remnants on centrophilic orbits \citep{Khan2012,Gualandris2017}.

AGN episodes provide an additional channel, both in low- and high-mass galaxies. In their outer regions, AGN discs can become self-gravitating and fragment, triggering in-situ star formation \citep{Goodman2003,Levin2007,Collin2008} favoring a top-heavy initial mass function \citep{Levin2007,Collin2008} and thereby producing a new population of compact remnants after stellar evolution. At smaller, sub-pc scales, interactions with the gaseous disc can capture pre-existing compact objects from the surrounding nuclear population, (damping their inclinations and embedding  them into the disc \citep{Generozov2023}), which in turn migrate inward through the disc and undergo gas-assisted binary formation \citep{McKernan2020,Tagawa2020}.

The presence of an IMBH or a secondary SMBH in the scenario discussed here, is motivated within hierarchical structure of galaxy formation, since repeated major and minor mergers can deliver additional massive black holes to galactic nuclei and form massive BH binaries at sub-pc separations \citep{Conselice2009,Martin2021,Khan2016}.

Indeed, the scattering of stars by hard SMBH binaries is commonly invoked as a mechanism for producing central stellar mass deficits in massive early-type galaxies, through the so-called binary-scouring process \citep{Merritt2006,Rusli2013}.
The same stellar-scattering process can also displace stars onto wider orbits \citep{Merritt2006}, thereby contributing to the diffuse low-surface-brightness stellar envelope observed around galaxies \citep{Merritt2016,Gilhuly2022}.

The nanohertz-emitting subset of the massive binary BH  population will be increasingly constrained by longer-baseline PTAs \cite{Siemens2013,Chen2019}, while binaries that proceed towards coalescence may be probed by LISA \citep{Seoane2023}. Currently, the NANOGrav 15-yr spectrum is consistent with a cosmological population of high mass ratio, sub-pc separation SMBH binaries \citep{Agazie2023}. 
We note that, in the ejection scenario considered here, the scatterer (an IMBH or a secondary SMBH) need not still be present at the time of the eventual NSX merger.
Once it reaches sub-pc separations, the secondary SMBH may coalesce with the central SMBH on a timescale comparable to, or shorter than, the merger time of the ejected NSX system.

The present work is an exploratory study of the dynamical consequences of this configuration, rather than  a population-synthesis model for producing NSX binaries and secondary massive black holes in galactic nuclei.
We assume that a population of compact NSX binaries and at least  one IMBH or secondary SMBH can coexist within the gravitational sphere of influence of the central SMBH, and we quantify the resulting ejection, survival, merger offset, and rate properties.

The outline of this paper is organiased as follows. In the next Section, we analytically derive the
velocity gain of the NSX COM during the scattering event as well as
the survival criterion indicating whether or not the binary is ionised by the encounter.
In Section~\ref{sec:offset_calc}, we integrate the post-slingshot orbit of the surviving binaries in a host-galaxy potential including the central SMBH, stellar component, and dark-matter (DM) halo, to calculate the associated  merger offsets.
In Section~\ref{sec:rates}, we apply the previous scattering experiments to estimate  the corresponding ejection rates of the considered NSX types.
Finally, we discuss and summarise our results in Section~\ref{sec:summary}

\section{The ejection of NSX binaries}
\label{sec:approach}
\subsection{Basic considerations}

\begin{figure}
\centering
\includegraphics[width=\columnwidth]{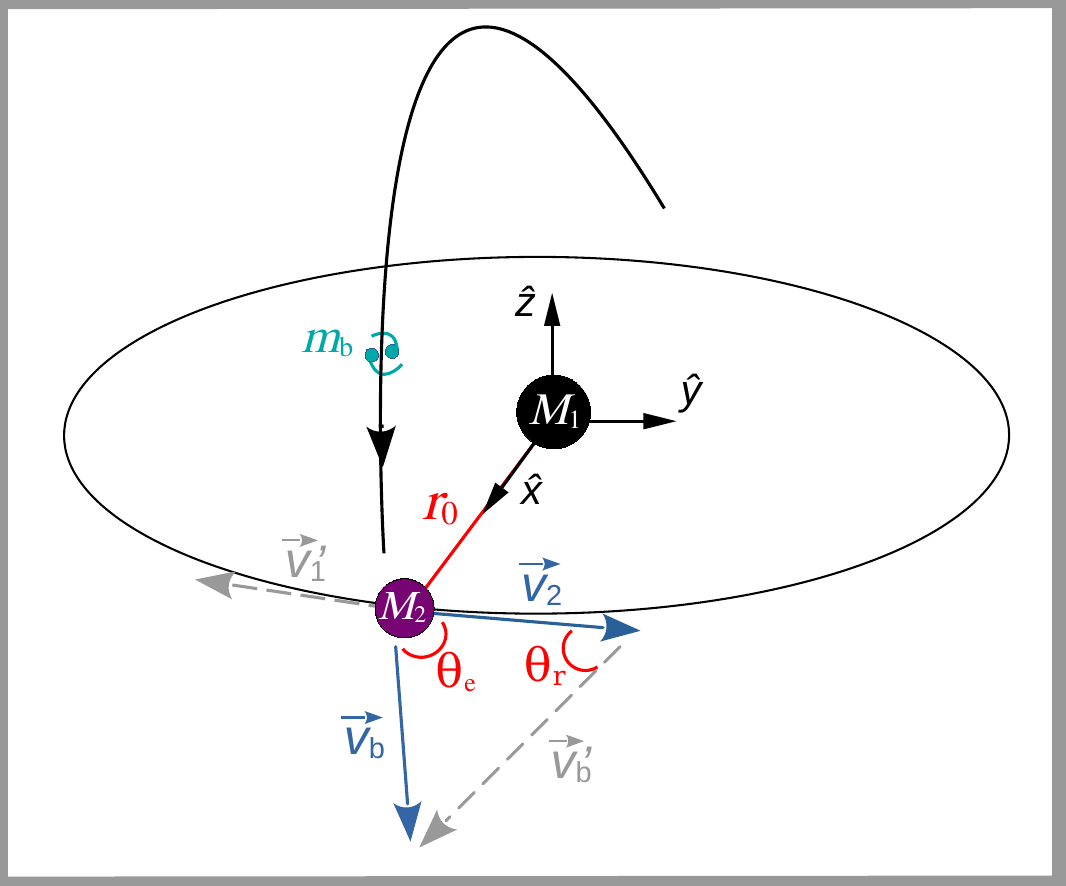}
\caption{
Schematic of a close encounter between an NSX binary of mass $m_\mathrm{b}$ and a BH of mass $M_2\gg m_\mathrm{b}$, both initially orbiting within the gravitational sphere of influence of the central SMBH of mass $M_1\gg M_2$.
Solid arrows show the velocity vectors of the NSX centre of mass (COM) and the BH $M_2$, respectively, in the frame where the primary SMBH is at rest, immediately before the NSX system enters the Hill sphere of $M_2$.
Dashed arrows indicate the corresponding velocity directions of the NSX COM and the primary SMBH in the co-moving frame of the BH $M_2$.
}
\label{fig:general_sk}
\end{figure}

We analyse the outcome of an IMBH or a secondary SMBH that encounters a population of NSX binaries within the gravitational sphere of influence of the central SMBH of mass $M_1$. This radius of influence is commonly defined as
\begin{equation}
 R_\mathrm{infl}= \frac{GM_1}{\sigma_*^2}
\label{R_infl}
\approx 3.0 \mathrm{pc} 
\left(\frac{M_1}{10^7\,\mathrm{M}_\odot}
\right)
\left(
\frac{120\, \mathrm{km}\,\mathrm{s}^{-1}}{\sigma_*}
\right)^2,
\end{equation}
where $G$ is the gravitational constant and $\sigma_*$ is the stellar velocity dispersion of the host galaxy bulge. The latter is often related to the central SMBH mass through the empirical $M_1-\sigma_*$ relation \citep[e.g.][]{Graham_2011}.

In what follows, we consider the NSX systems composed of an NS paired with either another NS, a stellar-mass BH, or a WD. 
The NSX system has a total mass $m_\mathrm{b}=m_1+m_2$, being $m_1$ and $m_2<m_1$ the masses of its components with separation $a_\mathrm{b}$ among them, such that the binary can potentially merge within a Hubble time.
The perturber that potentially scatters the trajectory of the NSX binaries has a mass $M_2$, and hereafter is referred to as  ``the BH $M_2$''. 
The following analysis considers that $m_\mathrm{b} \ll M_2 \ll M_1$, allowing the
NSX binaries and the BH $M_2$ to be treated as test particles under the gravitational potential of the central SMBH.
In addition, we consider encounters occurring at radial distances $r_0$ from the primary SMBH within 
$10^3R_\mathrm{g}<r_0<0.5R_\mathrm{infl}$, with $R_\mathrm{g}\equiv GM_1/c^2$.
This choice ensures that the trajectories of the perturber and of the NSX binaries can 
be described by Newtonian dynamics before and after their encounter.

The ejection of an NSX system critically depends on whether its encounter with the BH $M_2$ avoids disrupting the NSX binary while boosting its centre of mass (COM) above the  escape velocity in the frame where the primary SMBH is at rest, hereafter, ``the primary frame''.
We define this primary frame as the Cartesian system with unit vectors ($\hat{x}, \hat{y}, \hat{z}$), centred on the primary SMBH, as shown in Figure~\ref{fig:general_sk}.

Immediately before the BH $M_2$ significantly perturbs the trajectory of the NSX binary, the two systems move with velocities $\vec{v}_2$ and $\vec{v}_\mathrm{b}$ relative to the primary SMBH, respectively, forming an angle $\theta_\mathrm{e}$ with each other, as shown in Figure~\ref{fig:general_sk}.
The NSX trajectory is perturbed by the BH $M_2$ after entering the Hill sphere of $M_2$ hereafter, ``the Hill sphere'', whose radius $R_\mathrm{H}$ is related to the encounter radius $r_0$ as \citep{Murray_1999}:
\begin{equation}
\nonumber
\frac{R_\mathrm{H}}{r_0} =  \left(\frac{Q}{3} \right)^{1/3} 
\approx 
0.255\left(
\frac{Q}{0.05}
\right)^{1/3}
\,,
\label{R_H}
\end{equation}
with $Q\equiv M_2/M_1$.

\begin{figure}
\centering
\includegraphics[width=0.9\columnwidth]{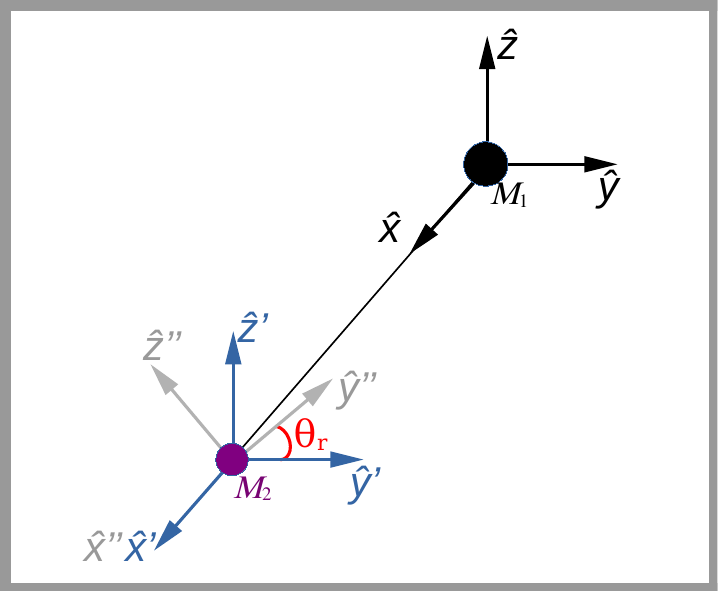}
\caption{
Schematic of the basis vectors of the three reference frames used in this analysis: $\hat{x}, \hat{y}$, and $\hat{z}$ for the frame centred on the primary SMBH; $\hat{x}^{\; \prime}, \hat{y}^{\; \prime}$, and $\hat{z}^{\; \prime}$ for the co-moving frame of the BH $M_2$; and $\hat{x}^{\; \prime\prime}, \hat{y}^{\; \prime\prime}$, and $\hat{z}^{\; \prime\prime}$ for the {\it rotated} $M_2-$frame, where $\hat{y}^{\; \prime\prime}$ is anti-aligned with the direction of the incoming NSX COM at the Hill-sphere boundary.
Both the primed and double-primed bases are centred on the BH $M_2$, and the $\hat{y}^{\; \prime\prime}$--$\hat{z}^{\; \prime\prime}$ plane is coplanar with the $\hat{y}^{\; \prime}$--$\hat{z}^{\; \prime}$ plane.}
\label{fig:3frames}
\end{figure}

During its passage through the Hill sphere, the COM of the NSX binary can be treated as a test particle moving under the gravitational influence of the BH $M_2$ alone. 
For convenience, we define the co-moving frame of the BH $M_2$ at the time of the encounter (hereafter ``the $M_2-$frame''), as the Cartesian system centred on the BH $M_2$ and with basis vectors $(\hat{x}^{\; \prime}, \, \hat{y}^{\; \prime},\, \hat{z}^{\; \prime})$ parallel to those of the primary frame, as illustrated in Figure~\ref{fig:3frames}.

\begin{figure}
\centering
\includegraphics[width=0.9\columnwidth]{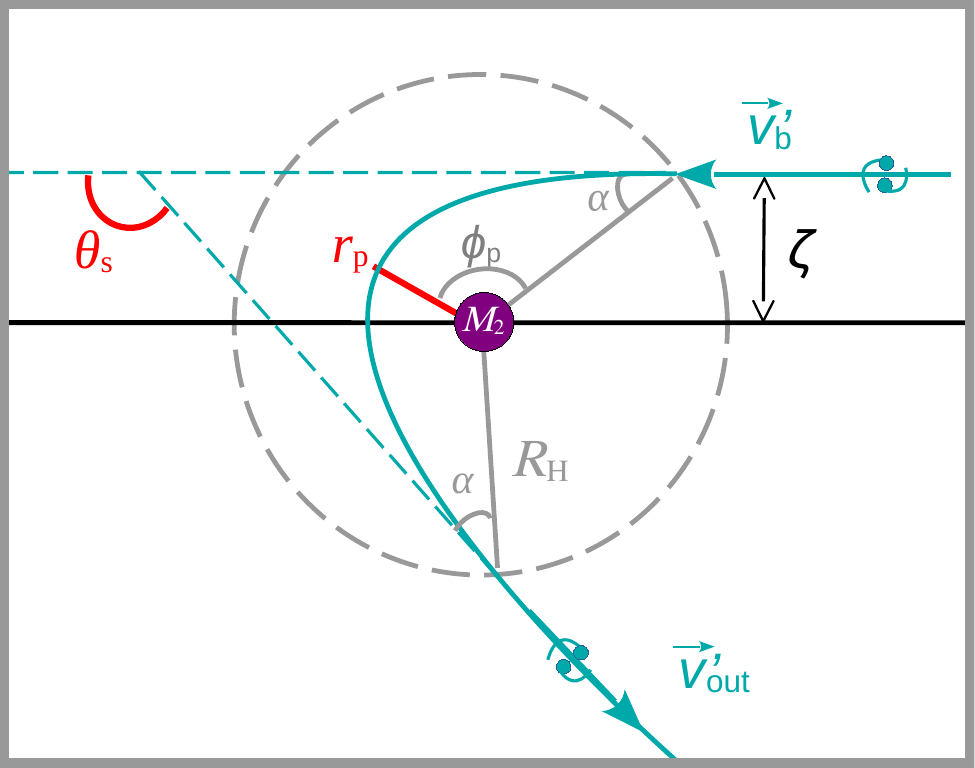}
\caption{
Two-dimensional schematic of the geometrical quantities associated with the orbit of the NSX COM during its passage through the Hill sphere of the BH $M_2$, in the co-moving frame of $M_2$.
The NSX orbit is represented by the solid hyperbola, while the Hill sphere is shown as a dashed circle.
}
\label{fig:sk_trig}
\end{figure}

In this $M_2-$frame, 
and at the boundary of the Hill sphere, the NSX COM has speed, energy, and angular momentum given by
\begin{align}
\label{v_b_sframe}
v^{\; \prime}_\mathrm{b} &=
\sqrt{
v_\mathrm{b}^2
+
v_\mathrm{2}^2  -
2v_\mathrm{b}v_\mathrm{2} \cos\theta_\mathrm{e} 
}\,,\\
\label{E_sframe}
E^{\; \prime} &= \frac{1}{2}m_\mathrm{b}v_\mathrm{b}^{\; \prime\,2} - \frac{G M_\mathrm{2}m_\mathrm{b}}{R_\mathrm{H}},\\
\label{L_sframe}
L^{\; \prime} &= m_\mathrm{b}v_\mathrm{b}^{\; \prime}\zeta,
\end{align}
respectively. Here, $v_\mathrm{b}=|\vec{v}_\mathrm{b}|$ and $v_2=|\vec{v}_2|$ are the speeds of the NSX COM and the BH $M_2$ measured in the primary frame, and $\zeta$ is the impact parameter of the encounter, as depicted in Figure~\ref{fig:sk_trig}.
In what follows, we use the notation $v_\mathrm{x}=|\vec{v}_\mathrm{x}|$ for the magnitude of any velocity vector.

\begin{figure}
\centering
\includegraphics[width=0.9\columnwidth]{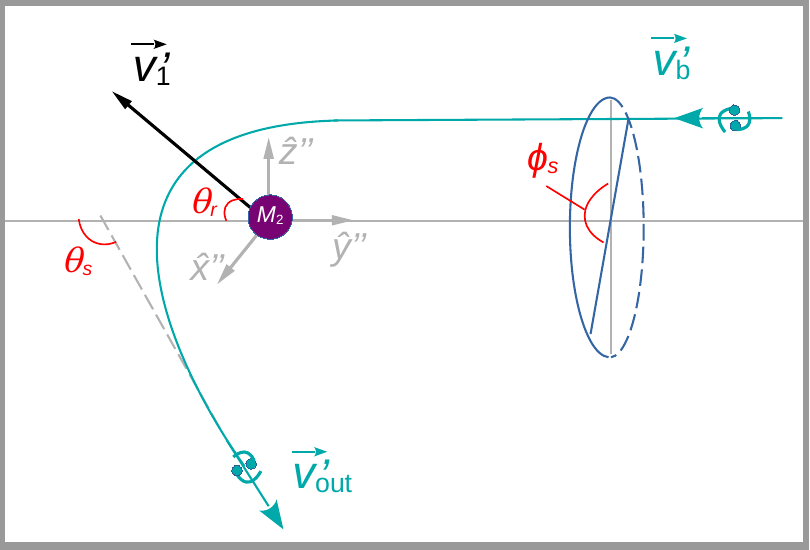}
\caption{
Three-dimensional schematic of the trajectory of the NSX COM in the frame co-moving with the BH $M_2$.
The double-primed Cartesian basis defines the rotated frame introduced in the text, in which the $\hat{y}^{\; \prime\prime}$ axis is anti-aligned with the incoming direction of the NSX system.
The closed, partially dashed curve marks a circular locus in the plane perpendicular to $\hat{y}^{\; \prime\prime}$, where the azimuthal angle $\phi_s$ is defined.
In this frame, the velocity of the primary SMBH represented by the vector $\vec{v}_1^{\; \prime}$, lies in the $\hat{y}^{\; \prime\prime}\,$--$\,\hat{z}^{\; \prime\prime}$ plane.
}
\label{fig:rotated_frame}
\end{figure}


Conservation of energy and angular momentum gives the eccentricity of the binary COM orbit in the $M_2-$frame as \cite{fetter2003theoretical}
\begin{equation}
e_\mathrm{cc} = \sqrt{1 + \left(1 - \frac{2R_\mathrm{B}}{R_\mathrm{H}}\right) \left(\frac{\zeta}{R_\mathrm{B}}\right)^2}\,\,,
\label{eccentricity}
\end{equation}
where
\begin{equation}
R_\mathrm{B} \equiv \frac{G M_\mathrm{2}}{v_\mathrm{b}^{\; \prime\,2}}.
\label{R_B}
\end{equation} 
The NSX system exits the Hill sphere when $e_\mathrm{cc}>1$, or equivalently when $2R_\mathrm{B}/R_\mathrm{H}<1$, corresponding to a hyperbolic orbit within the Hill sphere.

We introduce a simplified parametrisation for the relative speed $v^{\; \prime}_\mathrm{b}$ by assuming that, before the encounter, both the NSX system and the BH $M_2$ move with Keplerian speed $\sqrt{G M_1/r_0}$. This gives
 \begin{equation}
 v^{\; \prime }_\mathrm{b} = 
 \sqrt{2\left(1- \cos\theta_\mathrm{e}\right) GM_1/r_0}\,.
 \label{v_b_pr}
 \end{equation}
Under this approximation, we see that
encounters with angles $\theta_\mathrm{e}\gtrsim \pi/3$ result in NSX binary that exit the Hill sphere, since:
\begin{align}
\frac{2R_\mathrm{B}}{R_\mathrm{H}}&=
\frac{3^{1/3} Q^{2/3}}{1-\cos\theta_\mathrm{e}}
 \approx 0.621
\left( \frac{0.5}{1-\cos\theta_\mathrm{e}}\right)
\left(
\frac{Q}{0.1}
\right)^{2/3}.
\label{R_B_R_H}
\end{align}
This $e_\mathrm{cc}>1$ regime is the focus of the analysis in the following subsection.

\subsection{Velocity gain of the binary COM}

The encounter between the NSX binary and the BH $M_2$ can potentially increase the kinetic energy of the former relative to the primary SMBH through the gravitational deflection of its original trajectory by the BH $M_2$. This energy transfer is the basic mechanism underlying the ``slingshot effect'' \cite{Minovitch_1961,Flandro_1966,Sesana2006}.

To obtain the velocity gain of the NSX system in the present scenario, we first note that, in the $M_2-$frame, the binary exits the Hill sphere with speed $v^{\; \prime }_\mathrm{out} = v^{\; \prime }_\mathrm{b}$ and at an angle
\begin{equation}
\theta_\mathrm{s} = 2 \left[ \arcsin\left( \frac{\zeta}{R_\mathrm{H}} \right) + \phi_\mathrm{p} \right] - \pi,
\label{theta_h}
\end{equation}
relative to its entry direction. This relation follows directly from the geometry shown in Figure~\ref{fig:sk_trig}.
In equation~(\ref{theta_h}), $\phi_\mathrm{p}$ is the angle subtended between the direction of the binary periapsis and the direction of the binary entry point, as illustrated in Figure~\ref{fig:sk_trig}. From the classical central-force orbit equation \citep{fetter2003theoretical}, this angle is given by
\begin{equation}
\cos(\phi_\mathrm{p}) = \frac{1}{e_\mathrm{cc}} \left[ \left(\frac{R_\mathrm{B}}{R_\mathrm{H}}\right)\left(\frac{\zeta}{R_\mathrm{B}}\right)^2 - 1 \right].
\label{phi_p}
\end{equation}
In the limit $R_\mathrm{H}\rightarrow \infty$, and using the identity $\cos x=\pm1/\sqrt{1+\tan^2x}$, equation~(\ref{theta_h}) recovers the relation
$\tan(\theta_\mathrm{s}/2)=G M_2/(\zeta v_\mathrm{b}^{\; \prime \, 2})$, corresponding to the standard deflection angle of classical hyperbolic motion \cite{BinneyTremaine2008}.

In the frame of the primary SMBH, the velocity of the NSX binary upon exiting the Hill sphere is given by $\vec{v}_\mathrm{out} = \vec{v}^{\,\prime}_\mathrm{out} - \vec{v}^{\,\prime}_1$. Here, $\vec{v}^{\,\prime}_1$ is the velocity of the primary SMBH as seen in the $M_2-$frame. This vector forms an angle $\theta_\mathrm{r}$ with the direction of $\vec{v}_\mathrm{b}^{\; \prime }$, given by
\begin{align}
\label{sin_th_c}
\sin \theta_\mathrm{r}& = \frac{\sqrt{2}}{2} \sin\theta_\mathrm{e} 
(1-\cos\theta_\mathrm{e})^{-1/2}, \\
\label{cosin_th_c}
\cos\theta_\mathrm{r}& = 
\frac{\sqrt{2}}{2}\left(1-\cos\theta_\mathrm{e}\right)^{1/2}.
\end{align}
as follows from Figure~\ref{fig:general_sk}.

The magnitude of $\vec{v}_\mathrm{out}$ is then
\begin{equation}
v_\mathrm{out} = 
\sqrt{
v_\mathrm{out}^{\; \prime \, 2} + v_1^{\; \prime \, 2}
-2v^{\; \prime }_\mathrm{out}v^{\; \prime }_1\cos\theta_1
}\,,
\label{v_out_general}
\end{equation}
where $\theta_1$ is the angle between $\vec{v}^{\; \prime}_\mathrm{out}$ and $\vec{v}^{\; \prime}_1$.
The change in the NSX speed therefore depends critically on $\theta_1$, which encodes the slingshot geometry.

We compute $\theta_1$ in terms of $\theta_\mathrm{e}$ and $\theta_\mathrm{s}$ by introducing a convenient {\it rotated} $M_2-$frame with basis ($\hat{x}^{\; \prime\prime},\hat{y}^{\; \prime\prime},\hat{z}^{\; \prime\prime}$). This frame is obtained by rotating the $\hat{y}^{\; \prime }-\hat{z}^{\; \prime }$ plane around the $\hat{x}^{\; \prime }$ axis by the angle $\theta_r$, as illustrated in Figure~\ref{fig:3frames}.
In this rotated $M_2-$frame, the direction of $\hat{y}^{\; \prime\prime}$ is anti-aligned with the incoming direction of the NSX system. Therefore, $\vec{v}^{\; \prime}_\mathrm{out}$ and $\vec{v}^{\; \prime}_1$ can be written, in the rotated $M_2-$ frame,  as
\begin{equation}
\vec{v}^{\; \prime\prime}_\mathrm{out} =
\begin{pmatrix}
v^{\; \prime }_\mathrm{out} \sin\theta_\mathrm{s}\sin\phi_\mathrm{s}\\
-v^{\; \prime }_\mathrm{out} \cos\theta_\mathrm{s} \\
v^{\; \prime }_\mathrm{out}\sin\theta_\mathrm{s}\cos\phi_\mathrm{s}
\end{pmatrix},\,\,\,\,
\vec{v}^{\; \prime\prime}_1 =
\begin{pmatrix}
0 \\
-v_1^{\; \prime } \cos\theta_\mathrm{r} \\
v_1^{\; \prime } \sin \theta_\mathrm{r}
\end{pmatrix},
\label{ppvelocites}
\end{equation}
where $\phi_\mathrm{s}$ is the azimuthal angle around the incoming direction of the NSX binary, as sketched in Figure~\ref{fig:rotated_frame}.  
The angle $\theta_1$ is then obtained from
\begin{align}
\nonumber \cos\theta_1 =
\frac{\vec{v}^{\; \prime}_1\cdot \vec{v}^{\; \prime}_\mathrm{out}}{v_1^{\; \prime } v^{\; \prime }_\mathrm{out} } = &
\frac{ \vec{v}^{\; \prime\prime}_1 \cdot \vec{v}^{\; \prime\prime}_\mathrm{out}}{v_1^{\; \prime\prime} v^{\; \prime\prime}_\mathrm{out} }\\
=&
\cos\theta_\mathrm{r}\cos\theta_\mathrm{s}
+\sin\theta_\mathrm{r}\sin\theta_\mathrm{s}\cos\phi_\mathrm{s}.
\end{align}

Finally, using $|\vec{v}^{\; \prime}_1| = |\vec{v}_2| = \sqrt{GM_1/r_0}$, $v^{\; \prime }_\mathrm{out} = v^{\; \prime }_\mathrm{b}$, and combining equations~(\ref{v_b_pr}) and~(\ref{theta_h}--\ref{cosin_th_c}), the post-encounter speed of the NSX binary in the primary SMBH frame can be factorised by the local escape speed as
\begin{equation}
\label{v_out}
v_\mathrm{out} = 
F_\mathrm{esc} \sqrt{\frac{2GM_1}{r_0}}\,,
\end{equation}
with
\begin{equation}
\label{Fesc}
F_\mathrm{esc} =
\left[
\frac{1}{2}
+2(1-\cos\theta_\mathrm{e})\cos^2\Theta
+\sin\theta_\mathrm{e}\sin\left(2\Theta\right)\cos\phi_\mathrm{s}
\right]^{1/2}
\,,
\end{equation}

\begin{align}
&\Theta=\arcsin\left(\beta\right)
+
\arccos\left\{
\frac{\beta^2-\epsilon}{\sqrt{\epsilon^2+(1-2\epsilon)\beta^2}}
\right\},\\
& 
\beta\equiv \frac{\zeta}{R_\mathrm{H}}\,;\,\,
\epsilon\equiv\frac{R_\mathrm{B}}{R_\mathrm{H}}=
\frac{3^{1/3}Q^{2/3}}{2(1-\cos\theta_\mathrm{e})}\,.
\end{align}

The dimensionless function $F_\mathrm{esc}$ represents the output speed of the binary normalised by the local escape speed $\sqrt{2GM_1/r_0}$ from the central SMBH.
According to equation~(\ref{Fesc}), the binary can accelerate, decelerate, or approximately preserve its initial speed (here assumed to be the Keplerian speed $\sqrt{GM_1/r_0}$) after the encounter.
The binary speed is nearly unchanged when $\epsilon\ll \beta$ and $\beta\rightarrow 1$, namely when $\zeta\gg R_\mathrm{B}$ and $\zeta\rightarrow R_\mathrm{H}$. This limit gives $\Theta \rightarrow \pi/2$ and $F_\mathrm{esc}\rightarrow 1/\sqrt{2}$.
Conversely, when $\zeta \ll R_\mathrm{B}$, or equivalently $\beta\ll\epsilon$, and $\theta_\mathrm{e}\rightarrow\pi$, one obtains the extreme limit $\Theta \rightarrow \pi$ and $F_\mathrm{esc}\rightarrow 3/\sqrt{2}$. This case represents the maximum possible amplification of the binary speed, corresponding to a factor of $\sim 2$ above the local escape velocity from the potential of the central SMBH.

The speed boost predicted by equations~(\ref{v_out})--(\ref{Fesc}), however, is restricted by the requirement that the binary remain bound against the tidal field of the BH $M_2$.

\subsection{Binary survival}

A key condition for the ejection of NSX systems in the scattering events discussed in the previous section is that the binary survives tidal ionisation by the black hole $M_2$. The tidal perturbation is strongest when the binary COM reaches its closest approach to $M_2$, namely the periapsis distance $r_\mathrm{p}$ (see Figure~\ref{fig:sk_trig}),
\begin{equation}
r_\mathrm{p}
=
\sqrt{R_\mathrm{B}^2+\zeta^2}
-
R_\mathrm{B}.
\label{rp}
\end{equation}

The NSX binary is expected to be ionised when the energy injected by the tidal interaction becomes comparable to the magnitude of its binding energy,
\begin{equation}
E_\mathrm{b}
=
-\frac{G\mu m_\mathrm{b}}{2a_\mathrm{b}},
\end{equation}
where
\begin{equation}
\mu
=
\frac{m_1m_2}{m_1+m_2}
\end{equation}
is the reduced mass and $m_\mathrm{b}=m_1+m_2$ is the total mass of the binary.

In the tidal regime, $r_\mathrm{p}\gg a_\mathrm{b}$, the angular-averaged energy injection into the NSX binary by the BH $M_2$ can be estimated as derived in ~\ref{app:tidal_Delta_E}:
\begin{equation}
\left\langle \Delta E \right\rangle
=
\mu
\left(
\frac{GM_2 a_\mathrm{b}}{v_\mathrm{p} r_\mathrm{p}^2}
\right)^2,
\label{Delta_E_av}
\end{equation}
with
\begin{equation}
v_\mathrm{p}^2
=
v_\mathrm{b}^{\; \prime \, 2}
+
\frac{2GM_2}{r_\mathrm{p}}
\end{equation}
being the relative velocity between the NSX COM and $M_2$ at periapsis. Combining this expression with the binary binding energy gives
\begin{equation}
\frac{\langle\Delta E\rangle}{|E_\mathrm{b}|}
=
\frac{\left(
r_\mathrm{t}/r_\mathrm{p}
\right)^3}{\left[1+r_\mathrm{p}/(2R_\mathrm{B})\right]},
\label{ratE}
\end{equation}
where
\begin{equation}
r_\mathrm{t}
=
a_\mathrm{b}
\left(
\frac{M_2}{m_\mathrm{b}}
\right)^{1/3},
\end{equation}
is the nominal tidal radius of the NSX binary with respect to $M_2$, and $R_\mathrm{B}$ is given by equation~\ref{R_B}.

Thus, in this work, we classify as surviving systems those satisfying
\begin{equation}
\frac{\langle\Delta E\rangle}{|E_\mathrm{b}|}
<
0.25.
\end{equation}
Cases for which the tidal approximation $r_\mathrm{p}\gg a_\mathrm{b}$ is not achieved are considered as non-successful binary-ejection events in the parametric exploration below.

\subsection{Semi-major axis and eccentricity}

Tidal heating modifies the internal orbital energy of the NSX binary and therefore changes its semi-major axis. Comparing the binding energy before and after the encounter, and using the energy ratio defined above, the binary separation at ejection is
\begin{equation}
a_\mathrm{b,ej}
=
\frac{a_\mathrm{b}}
{1-\langle\Delta E\rangle/|E_\mathrm{b}|}.
\label{a_b_ej_main}
\end{equation}
Thus, for positive tidal heating, the binary becomes less bound and its semi-major axis increases. For our adopted survival threshold, this widening is smaller than about $33\%$.

The same tidal impulse can also excite the eccentricity of the internal binary orbit. Since we do not follow the internal phase of the NSX binary during the scattering event, we estimate this effect in an angular-averaged sense. Assuming that the pre-encounter internal orbit is nearly circular, that $a_\mathrm{b}/r_\mathrm{p}\ll1$, and that the instantaneous binary orientation at periapsis is isotropically distributed, the post-encounter eccentricity is estimated as the rms value derived in ~\ref{app:tidal_Delta_E}:
\begin{equation}
e_\mathrm{b,ej}
\approx
\sqrt{\left\langle e_\mathrm{b,ej}^2\right\rangle}
=
\left[
\frac{16}{5}
\frac{GM_2^2a_\mathrm{b}^3}
{m_\mathrm{b}v_\mathrm{p}^2r_\mathrm{p}^4}
\right]^{1/2}.
\label{e2_K_main}
\end{equation}
Equivalently, this can be written in terms of the effective energy ratio as
\begin{equation}
e_\mathrm{b,ej}
\approx
\left(
\frac{8}{5}
\frac{\langle\Delta E\rangle}{|E_\mathrm{b}|}
\right)^{1/2}.
\label{eb_DeltaE}
\end{equation}
For our adopted survival condition, equation~\ref{eb_DeltaE} implies $e_\mathrm{b,ej}\lesssim0.632$. Thus, in the tidal regime considered here, the eccentricity excitation of non-ionised binaries is expected to remain moderate.

\section{The merging offset}
\label{sec:offset_calc}

NSX binaries ejected from the sphere of influence of the primary SMBH can potentially reach the non-visible regions of the host galaxy and merge within a Hubble time.
The distance at which the NSX system merges, and hence the offset of the associated transient emission, is determined by both the merger time and the gravitational potential of the host galaxy.


The merger time $t_\mathrm{GW}$ due to gravitational-wave emission is governed by the coupled evolution of the binary semi-major axis and eccentricity \citep{Peters_1964}. Here, we estimate the post-encounter merger time of the NSX binary using the analytical prescription of Zwick {\it et al.} (2020) \cite{Zwick_2020}:,
\begin{equation}
 t_\mathrm{GW}
=
t_\mathrm{P} R_\mathrm{Z} Q_\mathrm{Z},
\label{t_GW}
\end{equation}
where
\begin{align}
&t_\mathrm{P}
=
\frac{5c^5(1+q)^2}{256G^3(m_1+m_2)^3q}
\frac{a_\mathrm{b,ej}^4}{f_e(e_\mathrm{b,ej})},
\label{t_P} \\
&R_\mathrm{Z}
=
8^{1-\sqrt{1-e_\mathrm{b,ej}}},\\
&Q_\mathrm{Z}
=
\exp\left[
\frac{5G(m_1+m_2)}
{c^2 a_\mathrm{b,ej}(1-e_\mathrm{b,ej})}
\right],\\
&f_e(e_\mathrm{b,ej})
=
\left(
1+\frac{73}{24}e_\mathrm{b,ej}^2+\frac{37}{96}e_\mathrm{b,ej}^4
\right)
(1-e_\mathrm{b,ej}^2)^{-7/2}.
\end{align}
Here, $a_\mathrm{b,ej}$ and $e_\mathrm{b,ej}$ are the post-encounter semi-major axis and eccentricity of the NSX binary, given by equations~\ref{a_b_ej_main} and~\ref{e2_K_main}, respectively. The quantity $t_\mathrm{P}$ is the usual Peters time-scale \citep{Peters_1964}, while $R_\mathrm{Z}$ and $Q_\mathrm{Z}$ are correction factors introduced by \cite{Zwick_2020}. The factor $R_\mathrm{Z}$ improves the treatment of eccentricity evolution, whereas $Q_\mathrm{Z}$ accounts for post-Newtonian corrections in the relativistic, high-eccentricity regime.

We then investigate the merger offset by integrating the Newtonian equation of motion, $\ddot{\vec{x}} = -\nabla \Phi_\mathrm{g}$, for the trajectory $\vec{x}(t)$ of the NSX COM up to an integration time given by $t_\mathrm{GW}$. We adopt the following superposition of gravitational potentials:
\begin{equation}
\Phi_\mathrm{g} = \Phi_\mathrm{BH}+\Phi_*+\Phi_\mathrm{DM}.
\label{gpotential}
\end{equation}
In equation~\ref{gpotential}, the terms $\Phi_\mathrm{BH}$, $\Phi_*$, and $\Phi_\mathrm{DM}$ account for the gravitational influence of the primary SMBH, the stellar component, and the DM component of the host galaxy, respectively.

The potential of the central SMBH at radius $r$ is $\Phi_\mathrm{BH} = -GM_1/r$.
For the stellar component, we consider both bulge and disc contributions, $\Phi_* = \Phi_\mathrm{bulge}+\Phi_\mathrm{disc}$.
Given the mass $M_1$ of the central SMBH, we estimate the bulge mass $M_\mathrm{bulge}$ from the phenomenological correlation derived by \cite{Kormendy_2013}:
\begin{equation}
\frac{M_1}{10^9\mathrm{M}_\odot} = 
\alpha \left( \frac{M_\mathrm{bulge}}{10^{11} M_\odot}\right)^\beta, 
\label{M1-Mbulge}
\end{equation}
with $\alpha = 0.49$ and $\beta= 1.17$.
We then model the bulge potential as spherically symmetric, adopting a Hernquist profile \citep{Hernquist1990} normalised by $M_\mathrm{bulge}$.
For the stellar disc, we adopt an axisymmetric Miyamoto--Nagai potential \cite{MiyamotoNagai_1975}, normalised by a disc mass $M_\mathrm{disc}$ obtained by assuming a bulge-to-total stellar mass ratio $B/T\leq1$.
With this parametrisation, the disc and total stellar masses scale with the bulge mass as
\begin{align}
&M_\mathrm{disc} = \left(\frac{1}{B/T}-1\right)M_\mathrm{bulge},\\
&M_\mathrm{*} = \frac{1}{B/T}M_\mathrm{bulge},
\end{align}
respectively, where $M_\mathrm{bulge}$ is obtained from equation~\ref{M1-Mbulge} for a given value of $M_1$.

Using this parametrisation, we explore two characteristic host-galaxy types: elliptical-like and spiral-like systems.
For the elliptical-like case, we set $B/T=1$, representing galaxies whose stellar mass is dominated by the bulge.
For the spiral-like case, we consider a disc-dominated system by setting $B/T=0.25$.

Finally, for the DM potential $\Phi_\mathrm{DM}$, we adopt a Navarro--Frenk--White profile \cite{NFW1996}, normalised by a halo mass $M_\mathrm{DM}$ estimated from the phenomenological calibration derived by \cite{Girelli_2020} (their equation 6), given the total stellar mass $M_*$.

We refer the reader to the Appendix for the detailed expressions used to define the potential components in equation~\ref{gpotential}.

\subsection{Post-slingshot orbit initialisation}
\label{pss_init}

To integrate the equation of motion for the NSX COM under the gravitational potential described above, we adopt Cartesian coordinates and initialise the orbit using the slingshot velocity derived in the previous section.
We therefore set the initial position and velocity of the integration as
$\vec{x}_0 = r_0 \hat{x}$ and
$\vec{v}_0 = \vec{v}_2+\vec{v}^{\; \prime}_\mathrm{out}$, respectively, where
$\vec{v}_2 = \sqrt{GM_1/r_0}\,\, \hat{y}$ is the velocity of the BH $M_2$ at the time of the encounter with the NSX system (see Figure~\ref{fig:general_sk}), and
$\vec{v}^{\; \prime}_\mathrm{out}$ is the velocity of the binary COM after the encounter, as measured in the $M_2-$frame.

To obtain the Cartesian components of $\vec{v}^{\; \prime}_\mathrm{out}$ in the $M_2-$frame, we apply the vector rotation
\begin{equation}
\vec{v}^{\; \prime}_\mathrm{out} = 
\mathcal{R}(\theta_\mathrm{r}) \vec{v}^{\; \prime\prime}_\mathrm{out},
\end{equation}
where 
\begin{equation}
\mathcal{R}(\theta_\mathrm{r}) = 
\begin{pmatrix}
1 & 0 & 0 \\
0 & \cos\theta_\mathrm{r} & -\sin\theta_\mathrm{r} \\
0 & \sin\theta_\mathrm{r} & \cos\theta_\mathrm{r}
\end{pmatrix}
\end{equation}
is the rotation matrix by an angle $\theta_\mathrm{r}$ around the $x^{\; \prime\prime}$-axis, and $\vec{v}^{\; \prime\prime}_\mathrm{out}$ is the post-slingshot velocity of the NSX COM in the rotated $M_2-$frame, given by equation~(\ref{ppvelocites}) (see also Figure~\ref{fig:rotated_frame}).
Therefore, the Cartesian representation of the post-slingshot velocity, namely the initial velocity of the numerical integration in the $M_1-$frame, is
\begin{equation}
\vec{v}_0 = 
\sqrt{\frac{2GM_1}{r_0}}
\begin{pmatrix}
\sqrt{1-\cos\theta_\mathrm{e}} \sin\theta_\mathrm{s}\sin\phi_\mathrm{s}\\
\frac{\sqrt{2}}{2}
\left[
1-(1-\cos\theta_\mathrm{e})\cos\theta_\mathrm{s} - \sin\theta_\mathrm{e}\sin\theta_\mathrm{s}\cos\phi_\mathrm{s}
\right]
\\
\frac{\sqrt{2}}{2}
\left[
(1-\cos\theta_\mathrm{e})\sin\theta_\mathrm{s}\cos\phi_\mathrm{s} - \sin\theta_\mathrm{e}\cos\theta_\mathrm{s}
\right]
\end{pmatrix},
\label{fullv0}
\end{equation}
whose norm recovers the post-slingshot speed given by equation~\ref{v_out}.

In summary, the orbit integration for the COM of an NSX binary after an encounter with the BH $M_2$ is fully determined by specifying the parameters
$M_1, M_2, r_0, \zeta,\theta_\mathrm{e}, \phi_\mathrm{s}$,
$m_1, m_2$, and $a_\mathrm{b}$, which define the gravitational potential, initial position, initial velocity, and merger time, namely $\Phi_\mathrm{g}$, $\vec{x}_0$, $\vec{v}_0$, and $t_\mathrm{GW}$, as described above.

Based on this numerical setup, we probe the distribution of slingshot-induced offsets by sampling $N_\mathrm{ev}$ encounters and then tracking the orbits of those binaries that remain internally bound after the encounter and whose post-encounter COM speeds exceed the local escape speed.
The time evolution of the binary COM is obtained by numerically integrating the equations of motion with the \texttt{solve\_ivp} routine of the \texttt{SciPy} package \citep{Virtanen2020}, using the explicit eighth-order Runge--Kutta method \texttt{DOP853}.
The accuracy of the orbit integrations is assessed in \ref{app:accuracy}, where we show the fractional variations of the energy and angular momentum of individual orbits.

\subsection{Simulating scattering events}
\label{subsec:sampling}

To simulate offset mergers produced by NSX scattering events, we fix the central SMBH mass, $M_1$, and sample uniformingly the following parameter ranges:
\begin{itemize}
\item 
$\log Q\in[-3.0,\, -1.0]$
\item 
$\log\left( r_0/R_g\right)\in 
[\log(R_\mathrm{min, eff}/R_g),\,\, \log(R_\mathrm{max, eff}/R_g)]$
\item
$\tilde{m}_1\in [m_\mathrm{1,min},\,\, m_\mathrm{1,max}]$
\item
$\tilde{m}_2\in [m_\mathrm{2,min},\,\, m_\mathrm{2,max}]$
\item 
$\log( a_\mathrm{b})\in [\log(a_\mathrm{b,max}), \log(a_\mathrm{b,min})]$
\item 
$\theta_\mathrm{e}\in[0,\,\,\pi]$
\item 
$\phi_\mathrm{s}\in[0,\,\,2\pi]$
\item 
$\sigma_\zeta\in[\sigma_\mathrm{min},\,\,\sigma_\mathrm{max}]$,
\end{itemize}
from which we define $M_2 = QM_1$, $m_1=\max(\tilde{m}_1,\tilde{m}_2)$,
$m_2=\min(\tilde{m}_1,\tilde{m}_2)$,
and
$\zeta = \sqrt{\sigma_\zeta/\pi}$.

In the parameter ranges above, we set 
$R_\mathrm{min,eff} =10^3GM_1/c^2$ and $R_\mathrm{max, eff}=0.5R_\mathrm{infl}$ as the radial boundaries within which the scattering scenario is valid.
The minimum radius ensures that the orbit of the NSX COM, both before and after the encounter, can be described using Newtonian dynamics under the potential of the central SMBH.
The maximum radius ensures that the encounter occurs within the sphere of influence of the central SMBH.
We define this influence radius using equation~\ref{R_infl}, together with the $M-\sigma$ calibration derived by \cite{Graham_2011}:
\begin{equation}
\log \left(\frac{M_1}{M_\odot} \right)=
\alpha + \beta \log\left(\frac{\sigma_*}{\sigma_{*,0}}\right),
\label{M-sigma}
\end{equation}
with $\alpha = 8.13$, $\beta=5.13$, and $\sigma_\mathrm{*,0}=200$ km s$^{-1}$.
Thus, by combining equations~(\ref{R_infl}) and~(\ref{M-sigma}), we parametrise the sphere of influence of the central SMBH solely in terms of its mass $M_1$.

As specified above, the impact parameter is obtained by uniformly sampling the encounter cross-section within the range defined by
$\sigma_\mathrm{min/max} = \pi \zeta_\mathrm{min/max}^2$,
where we set $\zeta_\mathrm{min} = 10^3 GM_2/c^2$ and $\zeta_\mathrm{max} = R_\mathrm{H}$ for the minimum and maximum impact parameters.
The lower value ensures that Newtonian dynamics adequately describes the NSX orbit within the Hill sphere of the BH $M_2$, whereas the upper value ensures that the NSX trajectory can be perturbed by the BH $M_2$.

For sampling the internal binary separation, we set the lower bound to
$a_\mathrm{b,min}=0.5 R_\odot$, allowing for a modest extrapolation below the
tightest observed detached NS-NS system, PSR J1946+2052 \cite{Stovall2018}
($a_\mathrm{b}\approx 1.05 R_\odot$, inferred from Kepler's third law). The upper bound is set to
$a_\mathrm{b,max}=10 R_\odot$, 
avoiding a dominant contribution from binaries with merger times exceeding the
Hubble time. As indicated above, $a_\mathrm{b}$ is sampled uniformly in
logarithmic scale (equivalent to a distribution $\propto a_\mathrm{b}^{-1}$) to
avoid artificially biasing the sample toward the largest separations and hence
toward the longest coalescence times. This choice is also qualitatively
consistent with the expectation that wider post-supernova compact binaries are
more vulnerable to disruption by natal kicks.

Likewise, the radial location of the encounter, $r_0$, is sampled uniformly in logarithmic space, corresponding to an encounter distribution that decreases with radius as $\propto r_0^{-1}$. This distribution can be interpreted as a higher concentration of NSX systems toward the inner regions of the galactic core.
Similarly, the sampling of $Q$ corresponds to a probability distribution for $M_2$ that decreases as $\propto M_2^{-1}$.
The distributions of NSX systems and secondary-BH masses can in principle be investigated in more detail. Nevertheless, such an exploration is beyond the scope of the present work, and we adopt the $\propto Q^{-1}$ and $\propto r_0^{-1}$ distributions as a reasonable first approximation.

The lower and upper limits for $m_1$ and $m_2$ are set according to the NSX
binary type, for which we consider NS--NS, NS--WD, and NS--BH binaries. We
sample the masses of NSs, WDs, and BHs uniformly within the ranges
$[1.2,2.2]\,M_\odot$, $[0.4,1.2]\,M_\odot$, and $[4,30]\,M_\odot$,
respectively. The adopted NS interval brackets the mass range of most measured
neutron stars in compact binaries, while allowing for relatively massive NSs
\citep{Ozel2016,Alsing2018}. The WD interval focuses on CO and moderately
massive ONe WDs relevant for NS--WD systems, and is consistent with observed WD
mass distributions and NS--WD population-synthesis calculations
\citep{Tremblay2016,Toonen2018}. For BHs, the lower limit lies above the
expected NS/BH transition while allowing for low-mass stellar BHs near the
lower-mass-gap regime \citep{Farr2011}. The upper limit is chosen to enclose the
BH masses of the observed NS--BH mergers reported by the LVK Collaboration, and
to allow for moderately massive BHs expected in dynamical formation channels
\citep{Abbott2021,Sedda2020}\footnote{The dynamical channel can produce
NS--BH binaries with BH masses larger than the upper limit adopted here. However, for increasingly massive BH components the production
of an electromagnetic counterpart is expected to be suppressed in the
highest-mass tail \citep{Sedda2020}.}. For each realisation, we sample the parameter ranges listed
above while keeping the NSX binary type fixed.

We integrate the post-encounter orbit, as described in the previous subsection, for parameter configurations that simultaneously satisfy
\begin{align}
\label{ecc_cond}
&e_\mathrm{cc}>1.0\,,\\
\label{rp_cond}
&a_\mathrm{b}/r_\mathrm{p}<0.1\,,\\
\label{ratE_cond}
&\frac{\Delta E}{|E_\mathrm{b}|}<0.25\,,\\
\label{Fesc_cond}
&F_\mathrm{esc}>1.0\,,\\
&t_\mathrm{GW}<10^{10} \mathrm{yr}\,.
\label{tH_cond}
\end{align}
These conditions select internally bound NSX systems whose COM speed exceeds the local escape speed (relative to the central SMBH) after the encounter, and that merge within a Hubble time.
The physical and dimensionless quantities entering conditions~(\ref{ecc_cond})--(\ref{tH_cond}) are derived in equations~\ref{eccentricity}, \ref{rp}, \ref{ratE},(\ref{Fesc}), and (\ref{t_GW}).
We note that these conditions do not necessarily imply escape from the host galaxy as a whole.
However, since the escape speed within the influence sphere of the central SMBH can range from hundreds to thousands of km s$^{-1}$, binaries satisfying conditions~(\ref{ecc_cond})--(\ref{tH_cond}) can still travel far from the nuclear region and potentially merge beyond the main stellar body of the host galaxy, even if they remain bound to the stellar and dark-matter potentials of the host.
Thus, throughout this work we refer to events satisfying conditions~(\ref{ecc_cond})--(\ref{tH_cond}) as ``successful ejections''.

\begin{figure}
   \centering
   \includegraphics[width=\hsize]{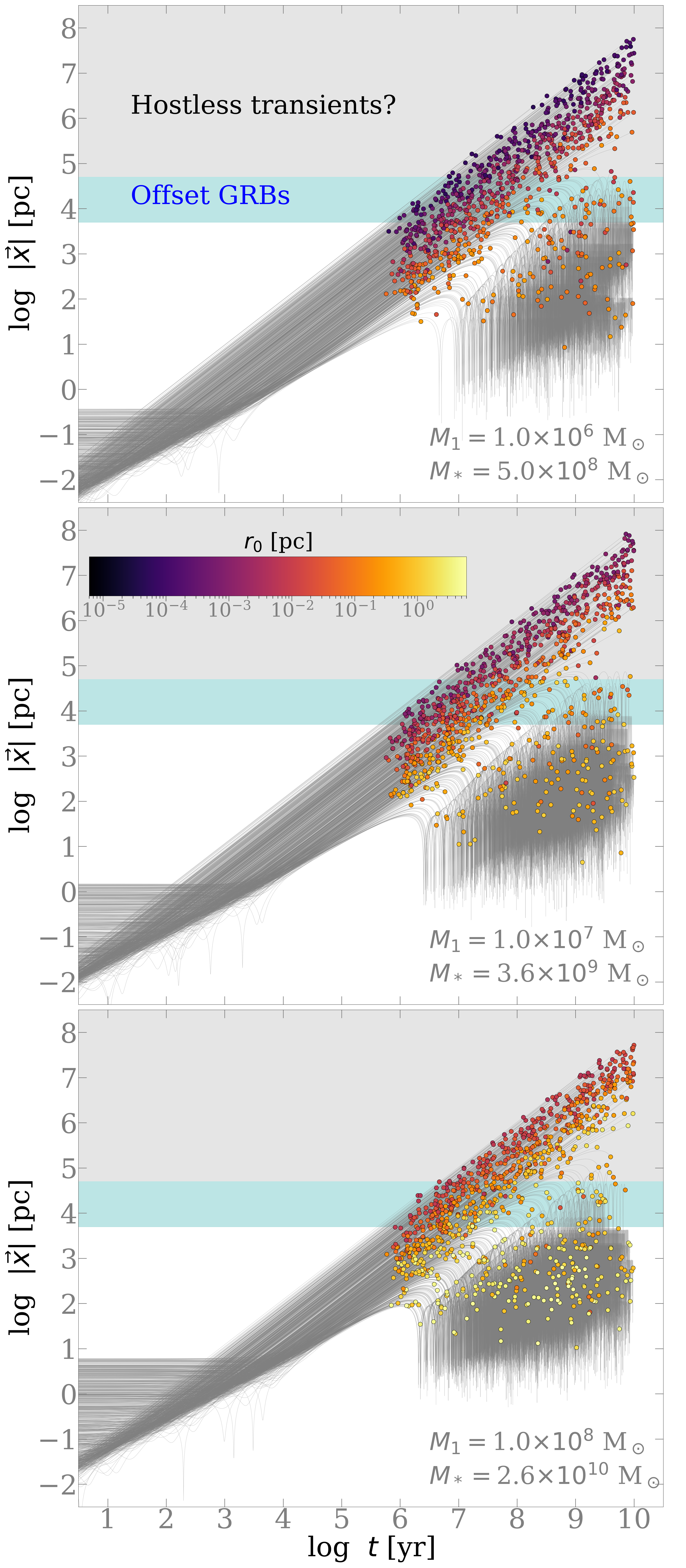}
\caption{
Simulated distance from the galaxy centre as a function of time for NS--NS binaries following a close encounter with a BH $M_2$ occurring within the gravitational sphere of influence of the central SMBH of mass $M_1$.
The curves are obtained by integrating the  equation of motion for the NSX COM in the elliptical-like galaxy potential, and correspond to sampled encounters that produce internally bound NS--NS binaries with COM speeds larger than the local escape speed at the scattering position (see text).
The coloured points at the end of each curve mark the binary merger locations, while the colour scale tracks the initial radial location, i.e. the encounter radius.
Different panels correspond to simulated ejections in galaxies with different stellar mass $M_*$, as indicated.
}
        \label{fig:d_t_M1s}
\end{figure}

Figure~\ref{fig:d_t_M1s} illustrates the outcome of the simulated scattering events for NS--NS binaries ejected from the nuclear region of the elliptical-like galaxy model.
The curves show the distance of the binary COM from the galaxy centre as a function of time.
The endpoint of each curve is marked with a coloured scatter point, indicating the binary merger location and therefore the offset of the associated transient emission.
The colour of these merger points tracks the radial location $r_0$ of the encounter (i.e., the initial radius of the integrated orbit). Each panel corresponds to a different assumed mass of the central SMBH, $M_1$, as labelled.

\begin{figure}
   \centering
   \includegraphics[width=\hsize]{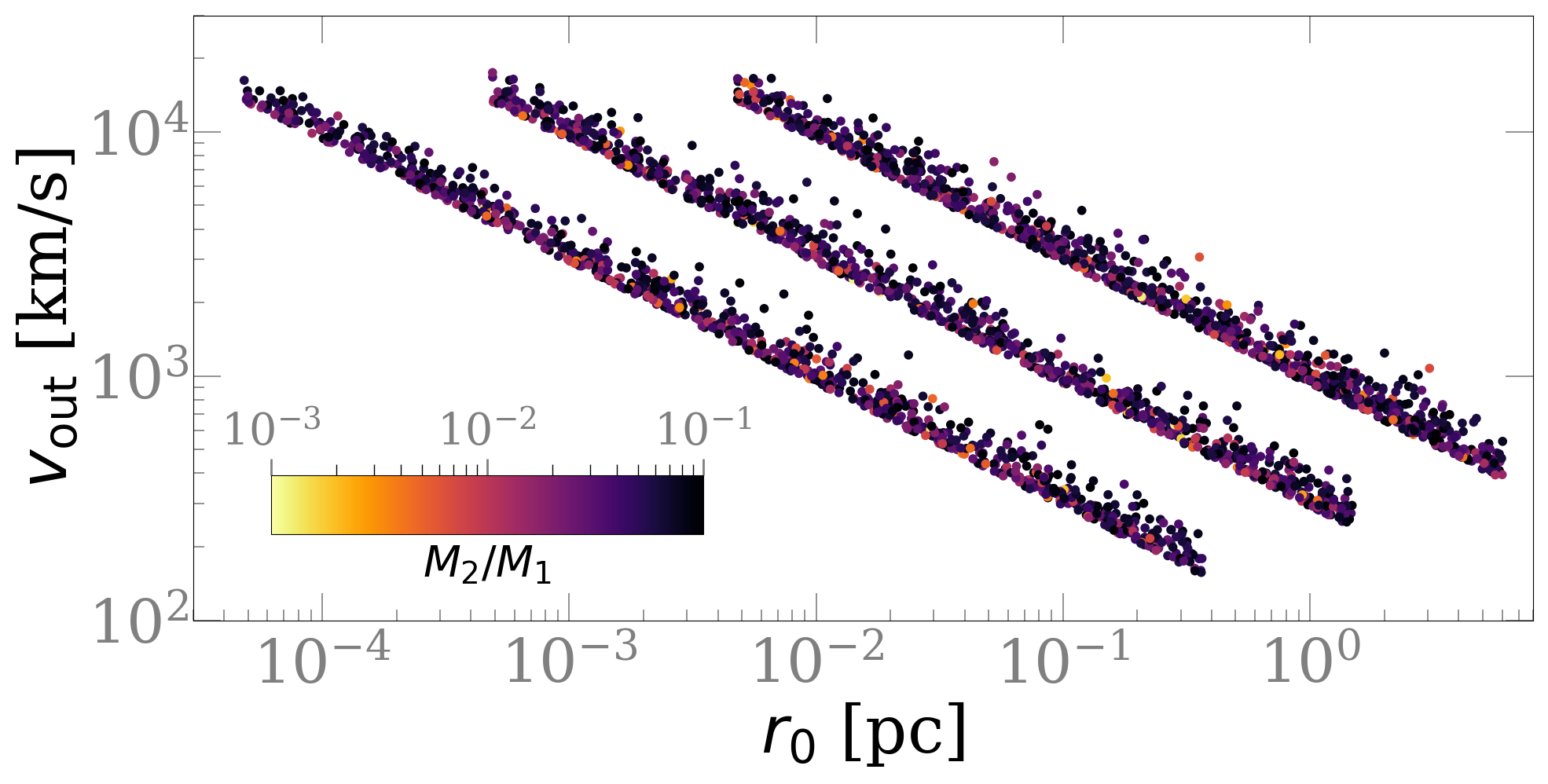}
\caption{
Initial speeds of the post-slingshot NS--NS binary orbits shown in Figure~\ref{fig:d_t_M1s} as a function of the encounter radius $r_0$.
The initial speed is obtained by sampling the model parameters as described in Section~\ref{subsec:sampling}, and corresponds to the post-slingshot speed given by equation~\ref{v_out}.
From left to right, the groups of scatter points correspond to the upper to lower panels in Figure~\ref{fig:d_t_M1s}, namely encounter events around central SMBHs of mass $M_1 = 10^6, 10^7$, and $10^8$ M$_\odot$, respectively.
The colour scale indicates the corresponding mass ratio $M_2/M_1$.
}
        \label{fig:v0s}
\end{figure}

Figure~\ref{fig:v0s} shows the initial speeds of the simulated orbits presented in Figure~\ref{fig:d_t_M1s} as a function of the radial location $r_0$ of the slingshot.
These initial speeds correspond to the post-slingshot speeds given by equation~\ref{v_out}.
Such encounters eject internally bound binaries with speeds ranging from hundreds to several thousands of km s$^{-1}$, with distributions piling up near the local escape speed.

As shown in Figure~\ref{fig:d_t_M1s}, the ejection mechanism proposed in this work can produce merger offsets ranging from kpc to Mpc scales within timescales shorter than $10^{10}$ yr, i.e. below a Hubble time.
In particular, NS--NS binaries must merge within tens to hundreds of Myr to produce offsets of tens of kpc, whereas Mpc-scale offsets are produced by binaries merging on Gyr timescales.
We observe that the correlation between merger offset and merger time remains similar for galaxies of different masses, which in the present framework are scaled by the mass of the central SMBH.
On the other hand, the central SMBH mass has a more pronounced impact on the encounter locations. In galaxies scaled to SMBHs with $M_1\sim 10^6$ M$_\odot$, extreme merger offsets require encounters occurring at $\sim 10^{-4}$--$10^{-2}$ pc,
whereas such extreme offsets are produced at $\sim 0.01$--$1$ pc when $M_1\sim 10^8$ M$_\odot$.

\begin{figure}
   \centering
   \includegraphics[width=\hsize]{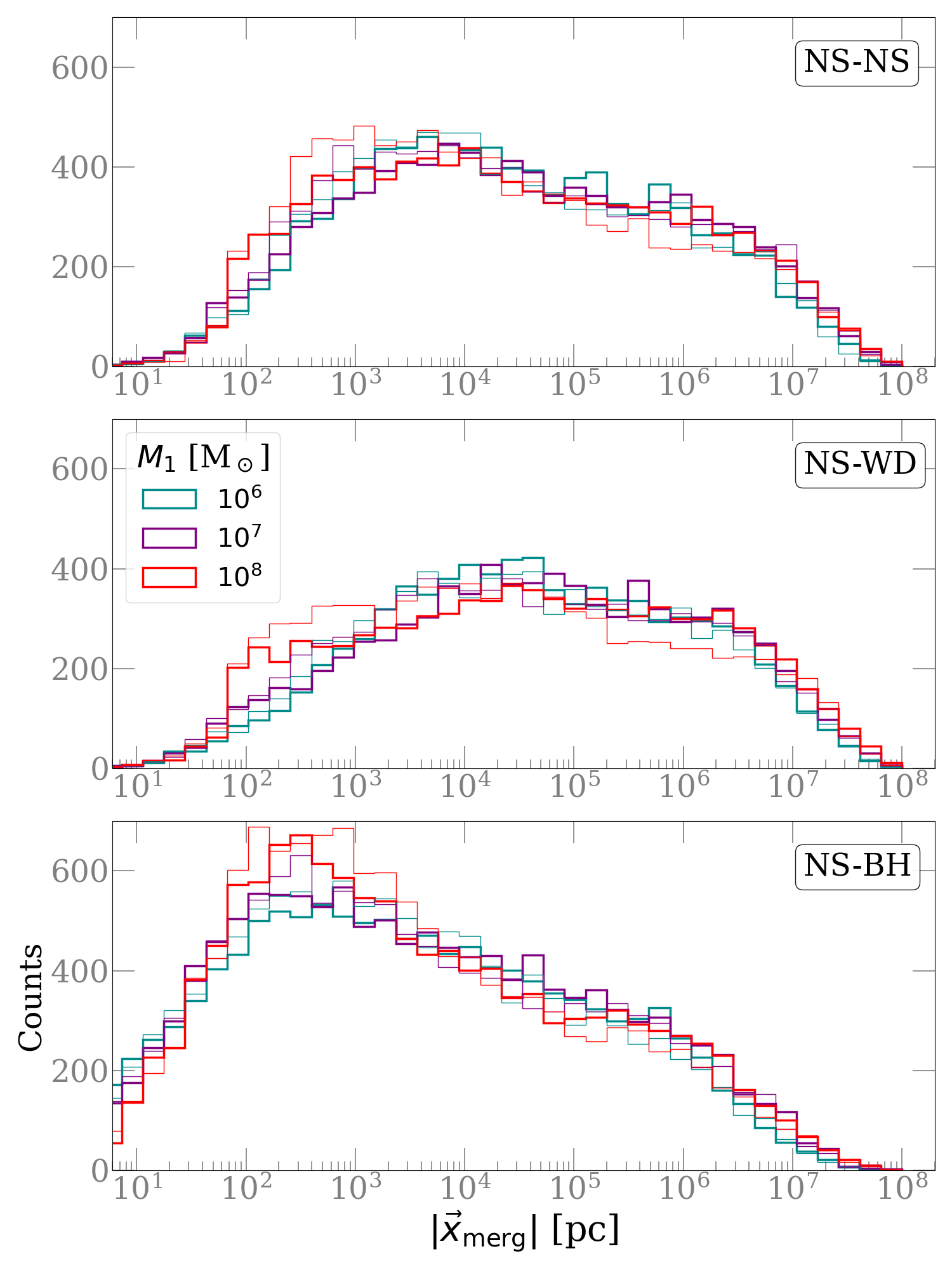}
\caption{
Distribution of merger offsets for slingshot-ejected NSX binaries in the galaxy potentials considered in this work. The distributions include only binaries that remain internally bound after the encounter and whose centre-of-mass speeds exceed the local escape speed, selected from $N_\mathrm{ev}=5\times10^5$ sampled encounters. The initial conditions for the subsequent orbit integrations are described in Section~\ref{pss_init}. Each panel corresponds to a fixed NSX binary type, as labelled. Colours indicate the central SMBH mass, $M_1$. Thick and thin histograms correspond to elliptical- and spiral-like galaxy potentials, respectively.
}
        \label{fig:histos}
\end{figure}

Figure~\ref{fig:histos} shows the distributions of merger offsets
where we include results for different NSX binary types and the two galaxy potential models described above.
The curves in each panel correspond to a
fixed NSX binary type, as indicated.
Curves of different colours correspond to galaxies with different SMBH masses $M_1$, as indicated, while thick and thin curves
correspond to the elliptical and spiral-like galaxy potential models, respectively.

The distributions are broad, but they exhibit a clear hierarchy with binary type. NS--WD systems typically merge at relatively large offsets, with
distributions peaking around \(\sim 10\)--\(100\,{\rm kpc}\) and with extended tails to larger distances.
NS--NS systems occupy an intermediate range, with the bulk of the distributions around \(\sim 1\)--\(10\,{\rm kpc}\). 
By contrast, NS--BH binaries preferentially merge at smaller offsets, with distributions peaking around \(\sim 0.1\)--\(1\,{\rm kpc}\), although with non-negligible tails extending to galactic and even intergalactic scales.

This hierarchy reflects the dependence of the gravitational-wave merger time on the binary masses
which scales approximately as
\(t_{\rm GW}\propto (m_1+m_2)^{-3}\) (see equation~\ref{t_P}). Therefore, the heavier NS--BH binaries merge more rapidly after ejection and have less time to travel away from the host nucleus. Conversely, the lower-mass NS--WD systems have longer inspiral times and can reach larger offsets before coalescence. The NS--NS case lies between these two limiting behaviours. The distributions also show that varying the central SMBH mass over \(M_1=10^6\)--\(10^8\,M_\odot\) produces only modest changes in the shape of the offset distribution compared with the much stronger dependence on the NSX binary type.


\begin{figure}
   \centering
   \includegraphics[width=\hsize]{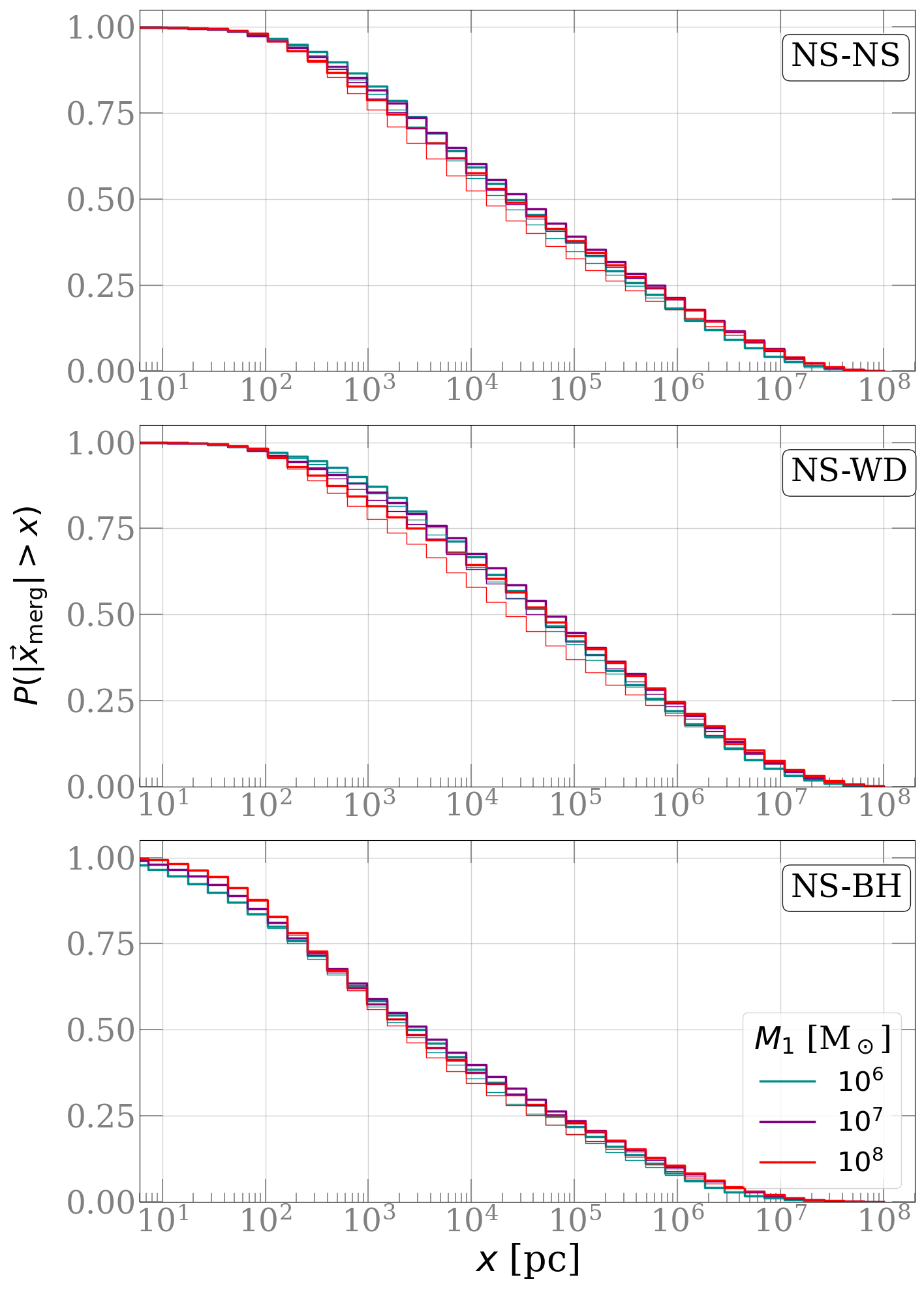}
\caption{
Complementary cumulative distribution functions (CCDFs) corresponding to the merger-offset distributions shown in Figure~\ref{fig:histos}. Each curve gives the probability $P(|\vec{x}_{\rm merg}|>x)$ that a binary merges at an offset greater than $x$ from the galactic centre. The panel, colour, and line-width conventions are the same as in Figure~\ref{fig:histos}.
}
        \label{fig:CCDFs}
\end{figure}

\begin{figure}
   \centering
   \includegraphics[width=\hsize]{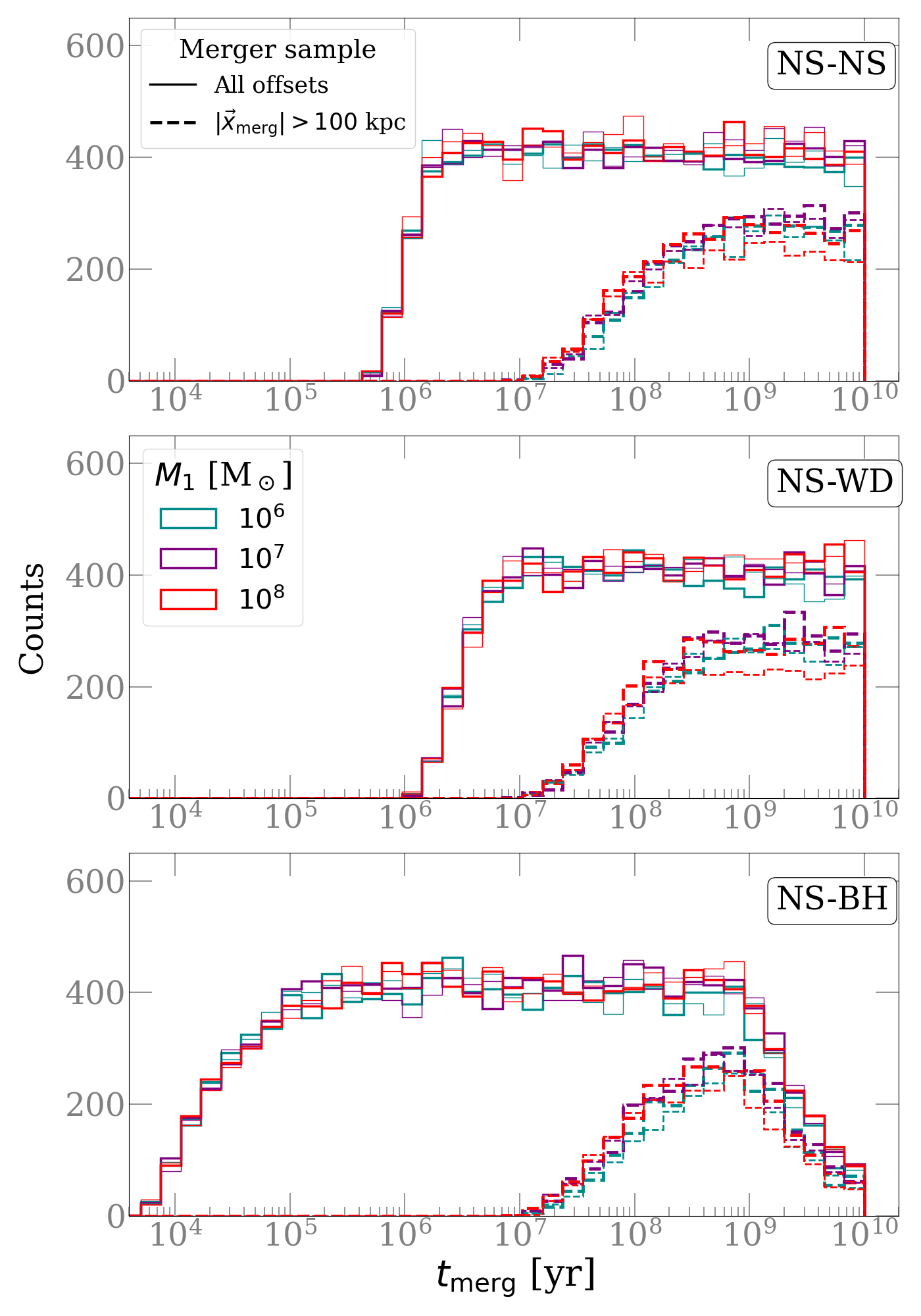}
\caption{
Distributions of merger times, $t_\mathrm{GW}$, for the NSX binaries drawn from the simulation outcomes presented in Figures~\ref{fig:histos} and~\ref{fig:CCDFs}. Solid curves include binaries at all merger offsets, whereas dashed curves show the subset that merges at galactocentric offsets $|\vec{x}_{\rm merg}|>100\,\mathrm{kpc}$. Panel labels, colours, and line widths follow the same conventions as in those figures.
}
        \label{fig:tm_gtr100}
\end{figure}

We note that the overall behaviour of the distributions corresponding to the elliptical and spiral-like galaxy models (thick and thin curves, respectively) is qualitatively similar.
This similarity indicates that the offset hierarchy is mainly set by the post-slignshot speed and the GW coalescence time, rather than by the detailed form of the stellar component of host potential. This latter modulates the trajectories, but does not erase the mass-dependent ordering among NS--WD, NS--NS, and NS--BH binaries.

Figure~\ref{fig:CCDFs} presents the complementary cumulative
distribution functions (CCDFs) corresponding to the merger-offset
distributions shown in Figure~\ref{fig:histos}.
These CCDFs quantifies the fraction of successfully
ejected binaries that merge beyond a given three-dimensional
galactocentric distance, $x$, under the adopted parameter sampling.
At 50 kpc, these fractions are approximately 25\% for NS--BH,
35--45\% for NS--NS, and 40--50\% for NS--WD systems.
At 100 kpc, the corresponding fractions are approximately
20\%, 30--40\%, and 35--45\%, respectively.
Thus, although NS--BH binaries typically merge closer to the
galactic centre, they retain a substantial tail of mergers at
large distances. Across the explored central SMBH masses and
host potentials, extreme merger offsets therefore occur in an
appreciable fraction of the successful realisations for all
three binary types. These fractions describe the simulated
three-dimensional merger distances; comparison with observed
host offsets additionally requires projection onto the sky.

Figure~\ref{fig:tm_gtr100} compares the distributions of merger times, $t_\mathrm{GW}$, for all NSX merger outcomes (solid curves) with those of the subset reaching three-dimensional merger offsets greater than $100\,\mathrm{kpc}$ (dashed curves). As expected, the full distributions extend to substantially shorter merger times, whereas selecting extreme offsets preferentially favours longer-lived systems, which remain unmerged long enough to travel farther from the galactic centre. Among the extreme-offset systems, NS--NS and NS--WD binaries merge predominantly on timescales of $0.1$--$10\,\mathrm{Gyr}$. By contrast, the corresponding NS--BH distributions are concentrated at shorter timescales, with most systems merging within $0.1$--$1\,\mathrm{Gyr}$.

We note that the merger-time distributions exhibit a sharp cutoff at $10^{10}\,\mathrm{yr}$, imposed by criterion~\ref{tH_cond}, which restricts successful ejections to binaries that merge within the adopted Hubble-time limit. This cutoff is more evident for the NS--NS and NS--WD systems, whereas the NS--BH distributions instead display a declining tail beginning at approximately $1\,\mathrm{Gyr}$. This decline, visible in both the full and extreme-offset NS--BH distributions, reflects the shorter gravitational-wave inspiral times of these more massive binaries. For illustration, adopting the largest binary separation sampled for all three binary classes, $a_{\mathrm{b,ej}}=10\,R_\odot$, together with $e_{\mathrm{b,ej}}=0.5$, equation~\ref{t_GW} gives $t_\mathrm{GW}\simeq0.87\,\mathrm{Gyr}$ for an NS--BH binary with component masses $1.4\,M_\odot$ and $20\,M_\odot$. For an NS--NS binary with component masses $1.4\,M_\odot$ and $1.4\,M_\odot$, the same separation and eccentricity yield $t_\mathrm{GW}\simeq83\,\mathrm{Gyr}$. Thus, the NS--BH distributions decline intrinsically before the Hubble-time selection limit is reached, whereas the NS--NS and NS--WD distributions extend towards longer inspiral times and are more visibly truncated at $t_\mathrm{GW}=10^{10}\,\mathrm{yr}$.

\section{The rate of binary ejections}
\label{sec:rates}

We use the scattering experiments described in the previous section to
estimate the averaged rate at which NSX binaries are ejected from
the nuclear region.

Each sampled parameter configuration describes a system containing one
secondary SMBH with mass
$M_{2,i}=Q_iM_1$, orbiting the central SMBH within a population of
$N_{\rm b, X}$ binaries
(X =NS-WD, NS-NS, NS-BH).
We approximate the NSX population as being
sufficiently well mixed within the radial domain of interest and define
its effective number density as
$
n_{\rm X}
=N_{\rm b,X} / V_{\rm eff},
$
where, for the spherical region adopted here,
\begin{equation}
V_{\rm eff}
=
\frac{4\pi}{3}
\left(
R_{\rm max,eff}^{3}
-
R_{\rm min,eff}^{3}
\right).
\label{eq:veff_rate}
\end{equation}

Among the parameter configurations sampled as described in
Section~\ref{subsec:sampling}, successful events are those that lie within
the adopted validity regime of the scattering model, preserve the
internally bound NSX binary, accelerate its centre of mass above the local
escape speed from the central SMBH, and result in a merger within the
adopted maximum time-scale. These requirements correspond to
conditions~(\ref{ecc_cond})--(\ref{tH_cond}).

Consider the $i$-th sampled parameter configuration. In the instantaneous
rest frame of the scatterer
(the secondary massive BH), the NSX COM approaches the perturber of mass $M_{2,i}$ with relative speed $v^{\; \prime}_{{\rm b},i}$, as defined in
equation~(\ref{v_b_pr}). For the values of $r_{0,i}$ and $Q_i$ in this
configuration, the impact parameter $\zeta_i$ is sampled uniformly in
cross-sectional area within an annulus of total area
\begin{equation}
\sigma_{{\rm enc},i}
=
\pi
\left(
\zeta_{{\rm max},i}^{2}
-
\zeta_{{\rm min},i}^{2}
\right),
\label{eq:sigma_enc_rate}
\end{equation}
where
\begin{equation}
\zeta_{{\rm min},i}
=
10^{3}\frac{GM_{2,i}}{c^{2}},
\qquad
\zeta_{{\rm max},i}
=
R_{{\rm H},i}.
\label{eq:zeta_limits_rate}
\end{equation}
Notice also that
$\zeta_{\rm min}$ is the minimum impact parameter relative to $M_2$ and
is distinct from the minimum galactocentric encounter radius,
$R_{\rm min,eff}=10^{3}GM_1/c^2$.

During a time interval $dt$, the relative motion covers a distance
$
d\ell_i
=
v^{\; \prime}_{{\rm b},i}\,dt.
$
The relative flow through the encounter cross-section therefore sweeps
an encounter tube with volume
\begin{equation}
dV_{{\rm sw},i}
=
\sigma_{{\rm enc},i}d\ell_i
=
\sigma_{{\rm enc},i}
v^{\; \prime}_{{\rm b},i}\,dt.
\label{eq:dvsw_rate}
\end{equation}
The expected number of NSX binaries contained within this volume is
\begin{equation}
dN_{{\rm enc},i}
=
n_{\rm X}dV_{{\rm sw},i}
=
n_{\rm X}
\sigma_{{\rm enc},i}
v^{\; \prime}_{{\rm b},i}\,dt.
\label{eq:dnenc_rate}
\end{equation}
The encounter rate associated with the $i$-th sampled parameter
configuration is consequently
\begin{equation}
\Gamma_{{\rm enc},i}
=
\frac{dN_{{\rm enc},i}}{dt}
=
n_{\rm X}
\sigma_{{\rm enc},i}
v^{\; \prime}_{{\rm b},i}.
\label{eq:gamma_enc_i}
\end{equation}

Both $\sigma_{\rm enc}$ and $v^{\; \prime}_{\rm b}$ vary among the scattering
experiments because they depend on the sampled parameters. We therefore
estimate the ensemble average of the encounter kernel
$\sigma_{\rm enc}v^{\; \prime}_{\rm b}$ using the $N_{\rm ev}$ sampled parameter
configurations. Because all realisations have equal statistical weight
under the sampling prescriptions described in
Section~\ref{subsec:sampling}, the corresponding Monte Carlo estimator is
\begin{equation}
\left\langle
\sigma_{\rm enc}v^{\; \prime}_{\rm b}
\right\rangle
\approx
\frac{1}{N_{\rm ev}}
\sum_{i=1}^{N_{\rm ev}}
\sigma_{{\rm enc},i}
v^{\; \prime}_{{\rm b},i}.
\label{eq:sigmav_average}
\end{equation}

The averaged encounter rate for a system containing one
secondary SMBH is therefore estimated as
\begin{equation}
\Gamma_{\rm enc,X}
\approx
n_{\rm X}
\left\langle
\sigma_{\rm enc}v^{\; \prime}_{\rm b}
\right\rangle
=
\frac{N_{\rm b, X}}
{V_{\rm eff}N_{\rm ev}}
\sum_{i=1}^{N_{\rm ev}}
\sigma_{{\rm enc},i}
v^{\; \prime}_{{\rm b},i}.
\label{eq:gamma_enc_rate}
\end{equation}

Only a subset of the sampled encounters produces a surviving and
successfully ejected NSX binary. We select these events using the
indicator function $S_i$, defined as $S_i=1$ when the $i$-th
sampled parameter configuration satisfies all conditions
(\ref{ecc_cond})--(\ref{tH_cond}), and $S_i=0$ otherwise.

The successful encounter kernel is therefore estimated as
\begin{equation}
\left\langle
\sigma_{\rm enc}v^{\; \prime}_{\rm b}
\right\rangle_{\rm succ}
\approx
\frac{1}{N_{\rm ev}}
\sum_{i=1}^{N_{\rm ev}}
\sigma_{{\rm enc},i}
v^{\; \prime}_{{\rm b},i}
S_i.
\label{eq:sigmav_success}
\end{equation}
The NSX ejection rate is consequently
\begin{equation}
\Gamma_{\rm ej,X}
\approx
\frac{N_{\rm b, X}}{V_{\rm eff}}
\left\langle
\sigma_{\rm enc}v^{\; \prime}_{\rm b}
\right\rangle_{\rm succ}
=
\frac{N_{\rm b, X}}
{V_{\rm eff}N_{\rm ev}}
\sum_{i=1}^{N_{\rm ev}}
\sigma_{{\rm enc},i}
v^{\; \prime}_{{\rm b},i}
S_i.
\label{gamma_ej_rate}
\end{equation}

In the present work, we do not attempt to derive the abundance of each NSX binary class in the inner sub-pc region. Such an estimate would require a dedicated population model accounting for stellar evolution, compact-object retention, dynamical processing, mass segregation, AGN-disc capture, and the merger history of the host galaxy, which is beyond the scope of this study. Instead, for the purpose of estimating the ejection rate, we adopt a common reference occupancy of \(N_{\mathrm{b},X}=30\) binaries for each NSX class (NS-WD, NS-NS, and NS-BH), independent of the SMBH mass \(M_1\). This choice should be interpreted as a phenomenological normalisation, not as a prediction for the relative abundance of the different NSX populations. For a nuclear star cluster containing \(\sim10^6\)--\(10^7\) stars \citep{Neumayer2020}, this corresponds to a small occupancy fraction of \(\sim3\times10^{-5}\)--\(3\times10^{-6}\) per binary class. Since the ejection rate scales linearly with \(N_{\mathrm{b},X}\), the rates obtained from equation~(\ref{gamma_ej_rate}) for \(N_{\mathrm{b},X}=30\) can be directly rescaled as
\begin{equation}
\Gamma_{\mathrm{ej},X}(N_{\mathrm{b},X}) =
\Gamma_{\mathrm{ej},X}^{(30)}
\left(\frac{N_{\mathrm{b},X}}{30}\right).
\end{equation}

\begin{figure}
   \centering
   \includegraphics[width=\hsize]{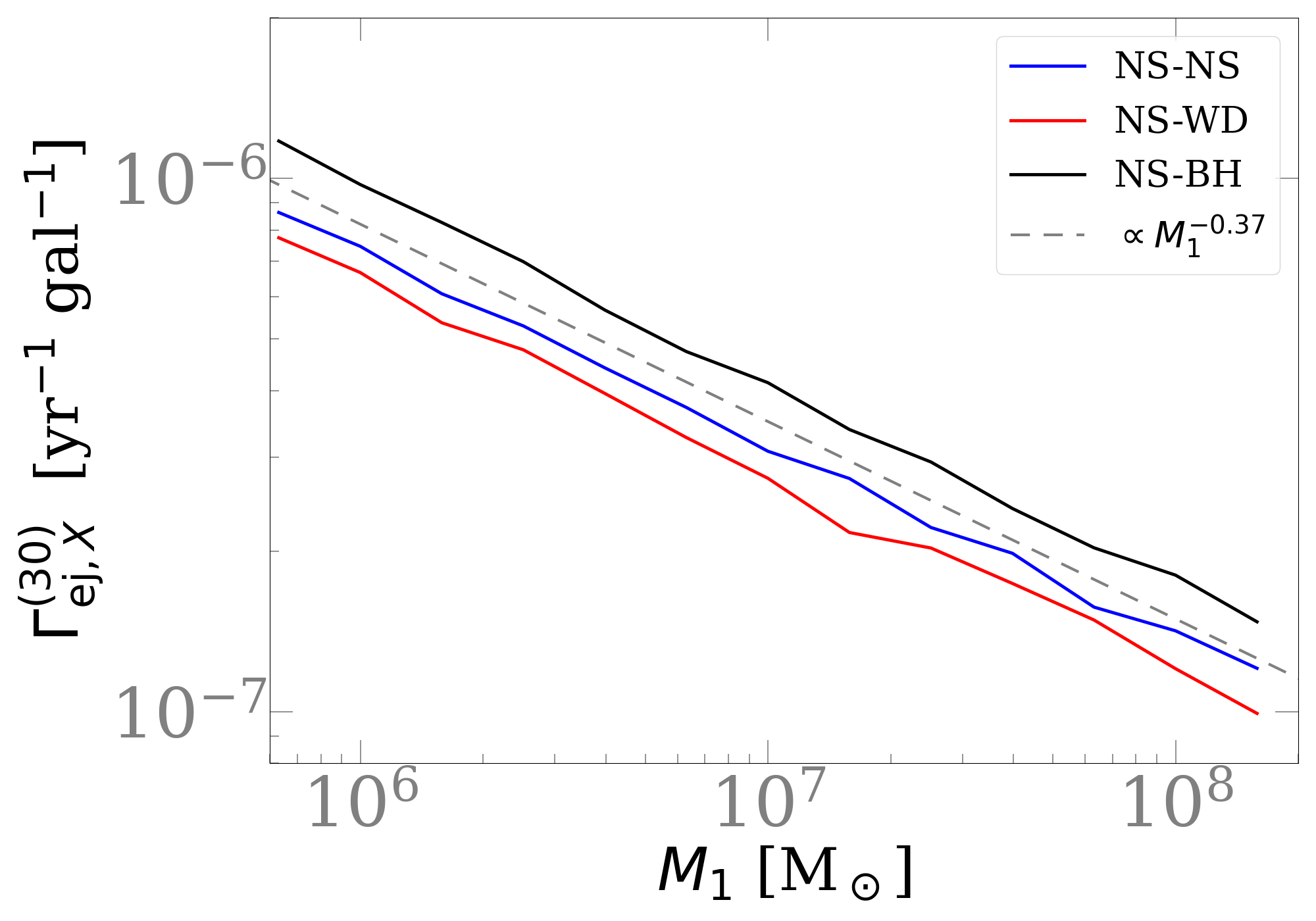}
\caption{
Ejection rates of successfully ejected NSX binaries as a function of the central SMBH mass $M_1$, for the three binary classes considered in this work: NS-NS, NS-WD, and NS-BH. The rates are estimated from Monte Carlo simulations of encounters between an NSX binary and a massive black hole within the influence radius of the primary central SMBH of the host galaxy. Successful ejections correspond to sampled encounters in which the NSX binary remains internally bound and its centre-of-mass velocity exceeds the local escape speed, satisfying the criteria defined by equations~\ref{ecc_cond}--\ref{tH_cond}. The solid curves are calculated from equation~\ref{gamma_ej_rate} using $N_\mathrm{ev}=5\times10^5$ sampled encounters for each binary type, and assume a reference occupancy of $N_{\mathrm{b},X}=30$ binaries of each class uniformly distributed within the SMBH influence radius. The rates can therefore be rescaled linearly for different choices of $N_{\mathrm{b},X}$. The dashed curve shows the approximate decreasing power-law trend, $\Gamma_\mathrm{ej}^{(30)}\propto M_1^{-0.37}$.
}
\label{fig:Grates}
\end{figure}

Figure~\ref{fig:Grates} shows the resulting reference rates \(\Gamma^{(30)}_{\mathrm{ej},X}\) as a function of the central SMBH mass \(M_1\), obtained from \(N_\mathrm{ev}=5\times10^5\) Monte Carlo encounters for each binary type. For \(M_1\sim10^6\)--\(10^7\,M_\odot\), the rates are of order \(\sim\mathrm{few}\times10^{-7}\)--\(10^{-6}\,{\rm yr^{-1}\,gal^{-1}}\), corresponding to roughly a few to ten ejections per \(10^7\) yr per galaxy for the adopted reference occupancy. In this fixed-\(N_{\mathrm{b},X}\) normalisation, the ejection rate decreases toward larger SMBH masses, approximately as \(\Gamma_{\mathrm{ej},X}^{(30)}\propto M_1^{-\alpha}\), with \(\alpha\approx0.37\) for the three NSX binary types considered. The relative ordering of the curves therefore reflects the different scattering and survival efficiencies of each binary class under the same assumed occupancy, rather than an assumed intrinsic abundance hierarchy.

\section{Summary and discussion}
\label{sec:summary}

In this paper, we have examined the gravitational scattering of compact binary systems within the influence radius $R_\mathrm{infl}$ of the central SMBH of galaxies, as an effective mechanism that produces offset transients.
This analysis is motivated by the growing number of GRBs and kilonova candidates observed at large projected offsets from their putative host galaxies,
and by the difficulty of explaining the most extreme offsets through the traditionally invoked natal-kick scenario alone \cite{Church2011, Alfradique2025, Bom2026}.

We analyse the scenario where
a population of compact binaries involving at least one neutron star (NSX binaries) has formed in the nuclear region of the host galaxy. Then,
the energy boost that ejects an NSX binary from this nuclear population is produced by the deflection of its centre-of-mass trajectory during an encounter with a much more  massive object.
We have focused on the case in which the perturber is either an IMBH or a secondary SMBH, since the large gravitational focusing radius of such objects can produce substantial deflections while still allowing the compact binary to remain internally bound.

To quantify this effect, we derived analytical conditions for
scattered binaries that escape the nuclear region as bound systems.
For analytical convenience, we parameterised the scattering encounters by assuming that both the perturber and the NSX binaries can be treated as test particles relative to the central SMBH, and that, prior to the encounters, they move with
non-relativistic speeds and Keplerian orbits around the primary SMBH. 
Within this framework, the survival
and escape conditions of a NSX binary are determined by the masses of the primary SMBH and the perturber, $M_1$ and $M_2$, the radial location of the encounter, $r_0$, the encounter angles and impact parameter, $\theta_\mathrm{e}$, $\phi_\mathrm{s}$, and $\zeta$, and the intrinsic properties of the NSX binary, namely $m_1$, $m_2$, and $a_\mathrm{b}$; see Section~\ref{sec:approach} for details.

To estimate the merger offsets of 
binaries successfully ejected from the nuclear region
(i.e. those satisfying conditions \ref{ecc_cond}-\ref{tH_cond}), we numerically integrated the equations of motion of their centers of mass up to their merger times. These integrations were performed in a multi-component gravitational potential including the central SMBH, the stellar component, and the DM halo of the host galaxy. The mass of the central SMBH, $M_1$, was used as the control parameter, while the stellar and DM components were scaled through empirically motivated prescriptions. This approach allowed us to simulate merger offsets in both elliptical-like and spiral-like galaxy models over a representative range of galaxy masses.


For the parameter space explored in this work (described in Section~\ref{subsec:sampling}), the distributions of the merger offsets
show a hierarchy with the NSX binary type. In both the elliptical-like and axisymmetric spiral-like host potentials, NS--WD binaries produce the largest offsets, with distributions peaking at tens to hundreds of kpc,
whereas NS--NS systems occupy an intermediate range, peaking at tens of kpc, and NS--BH binaries preferentially merge closer to the nucleus, around kpc scales, although with extended tails. This ordering 
is influenced by the GW coalescence time, which decreases as the mass of the neutron-star companion in the NSX binary increases. The weak dependence on the adopted host potential and on \(M_1\) over \(10^6\)--\(10^8\,M_\odot\) suggests that the large-scale galaxy model modulates the detailed trajectories but does not erase the offset hierarchy of the NSX type.

The CCDFs of the offset distributions shows that approximately 20\% of
successfully ejected NS--BH binaries, 30--40\% of NS--NS
binaries, and 35--45\% of NS--WD binaries merge at
three-dimensional galactocentric distances exceeding 100 kpc
(Figure~\ref{fig:CCDFs}). 
Then, using our numerical experiments, we derived a Monte Carlo estimate of a reference ejection rate by assuming the same fiducial occupancy, \(N_{\mathrm{b},X}=30\), for each NSX type. This choice
intends to provide a common normalisation under which the different scattering and survival efficiencies can be compared. For this reference occupancy, we find ejection rates of order 
\(\sim10^{-7}\)--\(10^{-6}\,{\rm yr^{-1}\,gal^{-1}}\).
This corresponds to roughly a few to ten successful ejections per \(10^7\) yr 
per galaxy hosting a secondary massive black hole.
Since \(\Gamma_{\mathrm{ej},X}\propto N_{\mathrm{b},X}\), these values can be linearly rescaled for any future estimate of the sub-pc NSX occupancy.
Given that a substantial fraction of
successful realisations lead to extreme merger distances,
these results motivates  
the proposed ejection mechanism as a candidate
channel for extreme-offset transients
\cite{Fong_2013, Fong2022, OConnor2022, Yang_2024Natur}.

The incidence of the required coexistence between
the massive black-hole pair and the nuclear NSX population remain
uncertain. Galaxy mergers motivate a plausible formation route
for the massive pair, but do not determine the abundance of the full
configuration. 
Our results therefore quantify the efficiency
and merger-offset outcomes of a candidate channel, rather than
its absolute cosmological event rate. Future work combining
nuclear compact-binary population models with the delivery and
orbital evolution of secondary massive black holes could
constrain its occurrence frequency and enable a quantitative comparison with the observed extreme-offset GRB population.

If the binary-ejection scenario discussed here operates in nature, 
the formation of compact binaries in the nuclear regions of galaxies can be probed with further detections of large offset transients.
At cosmological distances ($z\lesssim3$), the most detectable manifestations of this channel would likely be GRB events produced by relativistic jets viewed close to the line of sight, with projected offsets measured through X-ray and optical/near-infrared afterglow localisations combined with deep host-galaxy imaging \citep[e.g.][]{Fong2022,OConnor2022}. At smaller redshifts, however, the on-axis condition is not necessary: off-axis afterglows and kilonova-like optical/infrared counterparts could also reveal such mergers as large-offset or apparently hostless transients in wide-field time-domain surveys \citep[e.g.][]{Ivezic2019,Durgesh2026}.





\appendix
\section{Post-encounter binary heating, semi-major axis, and eccentricity}
\label{app:tidal_Delta_E}

In this appendix, we assess the energy injection $\Delta E$ into the internal binary orbit due to the tidal interaction with the BH $M_2$, as well as its effect on the binary semi-major axis and eccentricity.

Starting from the tidal impulse estimate, the energy injected into the internal binary orbit can be written as \citep{Gnedin_1999, Brown2018}:
\begin{equation}
\Delta E
\approx
\frac{1}{2}\mu \left| \Delta \vec{v}\right|^2,
\label{Delta_E}
\end{equation}
where
\begin{equation}
\mu=\frac{m_1m_2}{m_1+m_2},
\end{equation}
is the reduced mass of the binary, and
\begin{equation}
\Delta \vec{v}
=
\Delta \vec{\tau}\,t_\mathrm{p}.
\label{Delta_v}
\end{equation}
Here,
\begin{equation}
t_\mathrm{p}
=
\frac{r_\mathrm{p}}{v_\mathrm{p}},
\end{equation}
is the characteristic duration of the periapsis passage, with
\begin{equation}
v_\mathrm{p}^2
=
v^{\; \prime\,2}_\mathrm{b}
+
\frac{2GM_2}{r_\mathrm{p}},
\end{equation}
being the relative velocity between the NSX COM and $M_2$ at periapsis. The vector $\Delta \vec{\tau}$ denotes the tidal acceleration across the two components of the NSX binary.

To derive $\Delta \vec{\tau}$, and hence $\Delta E$, in terms of the parameters of our problem, we write the gravitational acceleration exerted by the BH $M_2$ at the position vector $\vec{r}$ as
\begin{equation}
\vec{g}(\vec{r})
=
-\frac{GM_2}{r^2}\hat{n}
=
-\frac{GM_2}{r^3}\vec{r}.
\end{equation}
Here, $\hat{n}=\vec{r}/r$ denotes the radial unit vector associated with $\vec{r}$, while its Cartesian components are denoted below by $n_i$. Let $\vec{r}_\mathrm{p}$ be the vector, of magnitude $r_\mathrm{p}$, pointing from the BH $M_2$ to the COM of the binary at periapsis, and let
\begin{equation}
\hat{n}_\mathrm{p}
\equiv
\frac{\vec{r}_\mathrm{p}}{r_\mathrm{p}}
\end{equation}
be the corresponding unit vector.

The tidal acceleration can then be written as the differential acceleration across the binary:
\begin{equation}
\Delta\vec{\tau}
=
\vec{g}(\vec{r}_\mathrm{p})
-
\vec{g}(\vec{r}_\mathrm{p}-\vec{a}_\mathrm{b}),
\label{Delta_tau}
\end{equation}
where $\vec{a}_\mathrm{b}$ is the instantaneous binary separation vector, with $|\vec{a}_\mathrm{b}|=a_\mathrm{b}$.

In the tidal regime, $a_\mathrm{b}/r_\mathrm{p}\ll1$, equation~\ref{Delta_tau} can be approximated by expanding $\vec{g}(\vec{r}_\mathrm{p}-\vec{a}_\mathrm{b})$ to first order in $a_\mathrm{b}/r_\mathrm{p}$. Thus, the components of the differential acceleration can be written as
\begin{equation}
\Delta \tau_i
\approx
\left(\frac{\partial g_i}{\partial x_j}\right)_{\vec{r}_\mathrm{p}}
a_{\mathrm{b},j},
\label{tidal_acc_i}
\end{equation}
where the index summation convention is adopted and $a_{\mathrm{b},j}$ is the $j$-th component of $\vec{a}_\mathrm{b}$. The quantity
\begin{equation}
\left(\frac{\partial g_i}{\partial x_j}\right)_{\vec{r}_\mathrm{p}}
=
\frac{GM_2}{r_\mathrm{p}^3}
\left(
3 n_{\mathrm{p},i}n_{\mathrm{p},j}
-
\delta_{ij}
\right)
\label{tidal_tensor}
\end{equation}
is the tidal tensor of the gravitational field of the BH $M_2$ evaluated at periapsis \citep[e.g.][]{Renaud2011,Duc2013}. In equation~\ref{tidal_tensor}, $n_{\mathrm{p},i}$ are the Cartesian components of the unit vector $\hat{n}_\mathrm{p}$, and $\delta_{ij}$ is the Kronecker delta.

Using equations~\ref{Delta_v} and~\ref{tidal_acc_i}, the tidal impulse can be written as
\begin{equation}
\Delta \vec{v}
=
\frac{GM_2}{v_\mathrm{p}r_\mathrm{p}^2}
\left[
3
\left(
\vec{a}_\mathrm{b}\cdot\hat{n}_\mathrm{p}
\right)
\hat{n}_\mathrm{p}
-
\vec{a}_\mathrm{b}
\right],
\label{Delta_v_vec}
\end{equation}
and the energy injected into the binary, equation~\ref{Delta_E}, becomes
\begin{equation}
\Delta E
=
\frac{\mu}{2}
\left(
\frac{GM_2 a_\mathrm{b}}{v_\mathrm{p} r_\mathrm{p}^2}
\right)^2
\left(
1+3\cos^2\psi
\right),
\end{equation}
where
\begin{equation}
\cos\psi
\equiv
\frac{
\vec{a}_\mathrm{b}\cdot\hat{n}_\mathrm{p}
}{
a_\mathrm{b}
}.
\end{equation}
For isotropically distributed binary orientations, $u=\cos\psi$ is uniformly distributed in the interval $[-1,1]$. Therefore,
\begin{equation}
\left\langle \cos^2\psi\right\rangle
=
\frac{1}{2}
\int_{-1}^{1}u^2\,du
=
\frac{1}{3}.
\label{cos2_av}
\end{equation}
In turn, the angular average of $\Delta E$ is
\begin{equation}
\left\langle \Delta E \right\rangle
=
\mu
\left(
\frac{GM_2 a_\mathrm{b}}{v_\mathrm{p} r_\mathrm{p}^2}
\right)^2.
\label{Delta_E_av}
\end{equation}

The NSX binary is expected to be ionised when the energy injected by the tidal interaction, \(\Delta E\), becomes comparable to the magnitude of its binding energy,
\begin{equation}
E_\mathrm{b} =
-\frac{G\mu m_\mathrm{b}}{2a_\mathrm{b}}.
\end{equation}
Thus, combining the above expressions with the binary binding energy gives
\begin{equation}
\frac{\langle\Delta E\rangle}{|E_\mathrm{b}|}
=
\frac{\left(
r_\mathrm{t}/r_\mathrm{p}
\right)^3}{\left[1+r_\mathrm{p}/(2R_\mathrm{B})\right]},
\end{equation}
where
\begin{equation}
r_\mathrm{t}
=
a_\mathrm{b}
\left(
\frac{M_2}{m_\mathrm{b}}
\right)^{1/3}
\end{equation}
is the nominal tidal radius of the NSX binary with respect to $M_2$, 
$R_\mathrm{B}=GM_2/v_\mathrm{b}^{\; \prime\,2}$, and
$m_\mathrm{b}=m_1+m_2$ is the total mass of the NSX binary.

The tidal heating modifies the internal orbital energy of the NSX binary and therefore changes its semi-major axis. Comparing the binding energy of the system before and after the encounter, and using the angular-averaged energy injection defined above, the binary separation at ejection is
\begin{equation}
a_\mathrm{b,ej}
=
\frac{a_\mathrm{b}}
{1- \langle\Delta E\rangle/|E_\mathrm{b}|}.
\label{a_b_ej}
\end{equation}
Thus, for positive tidal heating, the binary becomes less bound and its semi-major axis increases. For our adopted survival threshold, $\langle\Delta E\rangle/|E_\mathrm{b}|<0.25$, this widening is smaller than about $33\%$.

The same tidal impulse may also excite the eccentricity of the internal binary orbit. In contrast to the semi-major axis, however, the eccentricity depends not only on the change in internal energy, but also on the change in the internal angular momentum of the binary. Since we do not follow the internal phase of the NSX binary during the scattering event, we estimate this effect in an angular-averaged sense, consistently with the calculation of $\langle\Delta E\rangle$ above. We assume that the pre-encounter internal orbit is nearly circular, that the perturbation is weak and impulsive, that $a_\mathrm{b}/r_\mathrm{p}\ll 1$, and that the instantaneous orientation of the binary at periapsis is isotropically distributed.

Let
\begin{equation}
\vec{a}_\mathrm{b}=a_\mathrm{b}\hat{u}
\end{equation}
be the instantaneous relative separation vector of the compact binary, where $\hat{u}$ is the radial direction of the internal binary orbit. For an initially circular orbit, the corresponding relative velocity can be written as
\begin{equation}
\vec{w}
=
w\hat{w},
\qquad
w
=
\left(\frac{Gm_\mathrm{b}}{a_\mathrm{b}}\right)^{1/2},
\end{equation}
where $\hat{w}$ is perpendicular to $\hat{u}$.

For a nearly circular binary, a small in-plane velocity impulse changes the eccentricity according to the linear response of the eccentricity vector to an impulsive perturbation \citep[Eq.~2 of][]{Collins2008}. Equivalently, this expression follows from the small-kick, no-mass-loss, in-plane limit of the standard post-kick orbital-element relations for a circular binary \citep[Eqs.~9--10 of][]{Kalogera_1996}. In our notation, this gives
\begin{equation}
e_\mathrm{b,ej}^2
\approx
\left(
\frac{\Delta v_\mathrm{r}}{w}
\right)^2
+
4
\left(
\frac{\Delta v_\mathrm{t}}{w}
\right)^2,
\label{e_impulse}
\end{equation}
where
\begin{equation}
\Delta v_\mathrm{r}
=
\Delta\vec{v}\cdot\hat{u},
\qquad
\Delta v_\mathrm{t}
=
\Delta\vec{v}\cdot\hat{w},
\label{Delta_vr_vt}
\end{equation}
are the radial and tangential projections of the tidal impulse onto the internal binary orbit, with $\hat{u}\cdot\hat{w}=0$. The radial component changes the internal energy without changing the internal angular momentum at first order, whereas the tangential component changes both the internal energy and the internal angular momentum.

Using equations~\ref{Delta_v_vec} and~\ref{Delta_vr_vt}, the components of the tidal impulse are
\begin{align}
\Delta v_\mathrm{r}
&=
\frac{G M_2 a_\mathrm{b}}{v_\mathrm{p} r_\mathrm{p}^2}
\left[
3(\hat{u}\cdot \hat{n}_\mathrm{p})^2-1
\right],\\
\Delta v_\mathrm{t}
&=
\frac{3G M_2 a_\mathrm{b}}{v_\mathrm{p} r_\mathrm{p}^2}
(\hat{u}\cdot\hat{n}_\mathrm{p})
(\hat{w}\cdot\hat{n}_\mathrm{p}).
\end{align}
Here, $\hat{u}\cdot\hat{n}_\mathrm{p}=\cos\psi$. Since $\hat{w}$ is perpendicular to $\hat{u}$, we can write
\begin{equation}
(\hat{u}\cdot\hat{n}_\mathrm{p})
(\hat{w}\cdot\hat{n}_\mathrm{p})
=
\cos\psi\sin\psi\cos\phi,
\end{equation}
with $\psi\in[0,\pi]$ and $\phi\in[0,2\pi]$. As before, we consider isotropically distributed binary orientations. 
Then, using equation~\ref{cos2_av} and
\begin{equation}
\langle \cos^4\psi\rangle = \frac{1}{2}\int_{-1}^{1}u^4\,du = \frac{1}{5},
\end{equation}
the angular averages of $\Delta v_\mathrm{r}^2$ and $\Delta v_\mathrm{t}^2$ are
\begin{align}
\left\langle \Delta v_\mathrm{r}^2\right\rangle
&=
\left(\frac{GM_2a_\mathrm{b}}{v_\mathrm{p}r_\mathrm{p}^2}\right)^2
\left(
9\left\langle\cos^4\psi\right\rangle
-
6\left\langle\cos^2\psi\right\rangle
+
1
\right)
=
\frac{4}{5}
\left(\frac{GM_2a_\mathrm{b}}{v_\mathrm{p}r_\mathrm{p}^2}\right)^2,\\
\left\langle \Delta v_\mathrm{t}^2\right\rangle
&=
\left(
\frac{3G M_2 a_\mathrm{b}}{v_\mathrm{p} r_\mathrm{p}^2}
\right)^2
\left\langle
\cos^2\psi\sin^2\psi\cos^2\phi
\right\rangle
=
\frac{3}{5}
\left(\frac{GM_2a_\mathrm{b}}{v_\mathrm{p}r_\mathrm{p}^2}\right)^2,
\end{align}
and  from equation~\ref{e_impulse}, we obtain
\begin{equation}
\left\langle e_\mathrm{b,ej}^2\right\rangle
\approx
\frac{
\left\langle \Delta v_\mathrm{r}^2\right\rangle
+
4\left\langle \Delta v_\mathrm{t}^2\right\rangle
}
{w^2}
=
\frac{16}{5}
\frac{GM_2^2a_\mathrm{b}^3}{m_\mathrm{b}v_\mathrm{p}^2r_\mathrm{p}^4}.
\label{e2_K}
\end{equation}
Finally, we estimate the post-encounter eccentricity of the binary, for non-ionising cases, as the rms value
\begin{equation}
e_\mathrm{b,ej}
=
\sqrt{\left\langle e^2_\mathrm{b,ej}\right\rangle}.
\end{equation}

The induced eccentricity above can be written in terms of the effective energy ratio, as
\begin{equation}
e_\mathrm{b,ej}
\approx
\left(
\frac{8}{5}
\frac{\langle\Delta E\rangle}{|E_\mathrm{b}|}
\right)^{1/2}.
\label{eb_DeltaE}
\end{equation}
Thus, for our adopted survival condition, $
\frac{\langle\Delta E\rangle}{|E_\mathrm{b}|}
<0.25$, equation~\ref{eb_DeltaE} implies
\begin{equation}
e_\mathrm{b,ej}
\lesssim
0.632.
\end{equation}
Thus, in the tidal regime explored here, the eccentricity excitation of non-ionised binaries is expected to be moderate.

\section{Gravitational potential models}
\label{app:grav_potential}

In this appendix, we list the analytical expressions for the gravitational-potential components used to construct the elliptical-like and spiral-like galaxy models adopted in this work.

The stellar bulge is modelled with a Hernquist potential \citep{Hernquist1990},
\begin{equation}
\label{Pot_hq}
\Phi_\mathrm{bulge}\left(r\right) = - \frac{G M_\mathrm{bulge}}{r + a_\mathrm{h}},
\end{equation}
where $M_\mathrm{bulge}$ is the bulge mass and $a_\mathrm{h}$ is the scale radius. We define $a_\mathrm{h} = 0.55,R_{50}$ following \cite{Abbott_2017}, where $R_{50}$ is the stellar half-mass radius. The latter is obtained as a function of the stellar mass from the empirical relation derived by \cite{Zhang2019} (their equation 5), with parameters set according to the bulge-to-total ratio, B/T, specified in their Table 1. We use B/T$=1.0$ and B/T$=0.25$ for the elliptical-like and spiral-like cases, respectively.

The DM component is modelled using a Navarro--Frenk--White (NFW) potential \citep{NFW1996},
\begin{equation}
\Phi_\mathrm{DM}\left(r\right) = -\frac{4\pi G \rho_s r_\mathrm{s}^3}{r}
\ln\left(1 + \frac{r}{r_\mathrm{s}}\right),
\end{equation}
where $\rho_s$ and $r_\mathrm{s}$ are the characteristic density and scale radius of the DM halo, respectively. These quantities are related to the halo mass $M_{200}$ through the concentration parameter $c_{200}$ as
\begin{equation}
M_{200} = 4\pi \rho_s r_s^3 \left[
\ln\left(1+c_{200}\right) - \frac{c_{200}}{1+c_{200}}
\right].
\label{Mh}
\end{equation}

For a galaxy with total stellar mass $M_*$, we determine $\rho_s$ and $r_s$ as follows. We first estimate the DM halo mass $M_{200}$ using the stellar-to-halo mass relation of \cite{Moster2010,Girelli_2020},
\begin{equation}
\frac{M_*}{M_{200}} =
2A(z)
\left[
\left(\frac{M_{200}}{M_A}\right)^{-\beta(z)}
+
\left(\frac{M_{200}}{M_A}\right)^{\gamma(z)}
\right]^{-1},
\end{equation}
adopting the values of $A$, $M_A$, $\beta$, and $\gamma$ given in Table 1 of \cite{Girelli_2020} for the redshift bin $0<z<0.2$.
In this first study, we adopt only the central stellar-to-halo mass relation and do not include its intrinsic scatter. We plan to incorporate this additional source of uncertainty in future work.

Given $M_{200}$, we compute the concentration parameter $c_{200}$ with the \texttt{Colossus} toolkit \citep{Diemer_2018}, adopting the ``ishiyama21'' mass--concentration relation \citep{Ishiyama2021}. The halo scale radius is then obtained as $r_\mathrm{s}=R_{200}/c_{200}$, where
\begin{equation}
R_{200} = \left[\frac{3 M_{200}}{4\pi \Delta_{200} \rho_\mathrm{c}}\right]^{1/3},
\end{equation}
with $\Delta_{200}=200$ and $\rho_\mathrm{c}=1.4\times10^{11},M_\odot,\mathrm{Mpc^{-3}}$ denoting the critical density of the Universe. Finally, given $M_{200}$, $c_{200}$, and $r_s$, the characteristic density $\rho_s$ is obtained from equation~\ref{Mh}.

For spiral-like hosts, in addition to the Hernquist bulge component in equation~\ref{Pot_hq}, we also include a stellar disc. 
To allow a simple and fast evaluation of the disc gravitational potential at arbitrary three-dimensional positions, we model it with the axisymmetric Miyamoto--Nagai potential \citep{MiyamotoNagai_1975}
\begin{equation}
\Phi_\mathrm{MN} =
-\frac{GM_\mathrm{disc}}{
\sqrt{
\varrho^2 + \left[ a_\mathrm{MN}+\sqrt{z^2 +  b_\mathrm{MN}^2}\right]^2
}
},
\end{equation}
where $\varrho = \sqrt{x^2+y^2}$ is the cylindrical radius, $M_\mathrm{disc}$ is the stellar disc mass, scaled as described in Section~\ref{sec:offset_calc}, and $a_\mathrm{MN}$ and $b_\mathrm{MN}$ are the radial and vertical scale lengths of the disc, respectively.
Following \cite{Courteau2007}, we set $a_\mathrm{MN}=R_{50}/1.678$, and we adopt $b_\mathrm{MN}=0.1a_\mathrm{MN}$ to define the flattening of the profile.
The half-mass radius $R_{50}$ is obtained as for the Hernquist component, using the empirical parametrisation of \cite{Zhang2019} for the total stellar mass, i.e. bulge plus disc mass, and the adopted bulge-to-total ratio.

\section{Numerical accuracy of orbit integration}
\label{app:accuracy}

In this appendix, we assess the numerical accuracy of the orbit integrations performed for the NSX centre of mass in the galactic potentials considered in this work. As a diagnostic, we monitor the fractional variations of the conserved quantities along each numerical trajectory, from the initial post-slingshot position up to the time at which the binary is predicted to merge.

For the elliptical-like host model, all potential components are spherically symmetric. Therefore, both the orbital energy and the magnitude of the total angular momentum vector must be conserved along the trajectory. For the spiral-like host model, the inclusion of the Miyamoto--Nagai disc makes the total potential axisymmetric rather than spherically symmetric. In this case, the orbital energy remains conserved, but only the $z$-component of the angular momentum is an exact integral of motion.

We define the fractional energy variation as
\begin{equation}
\delta E(t) =
\left|
\frac{E(t)-E_0}{E_0}
\right|,
\end{equation}
where $E_0$ is the initial orbital energy. For the elliptical-like case, we also compute
\begin{equation}
\delta L(t) =
\left|
\frac{L(t)-L_0}{L_0}
\right|,
\end{equation}
where $L=|\vec{L}|$ is the magnitude of the total angular momentum. For the spiral-like case, we instead monitor
\begin{equation}
\delta L_z(t) =
\left|
\frac{L_z(t)-L_{z,0}}{L_{z,0}}
\right|,
\end{equation}
since $L_z$ is the conserved angular-momentum component in an axisymmetric potential.

\begin{figure*}
   \centering
   \includegraphics[width=\hsize]{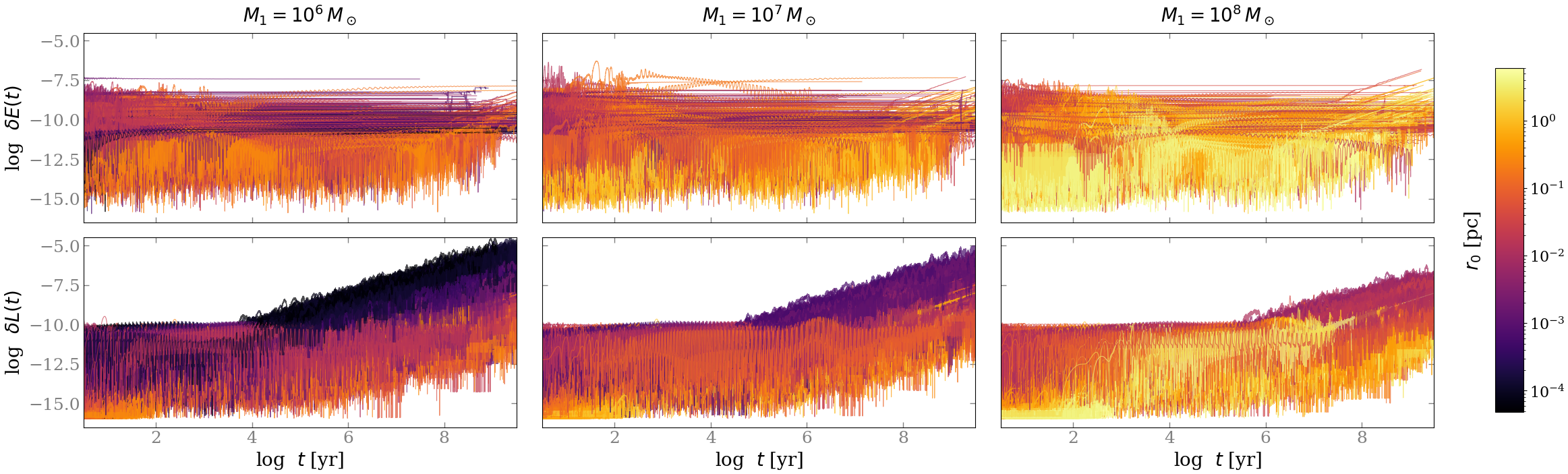}
\caption{
Time evolution of the fractional variations of the orbital energy and angular-momentum magnitude for $5\times10^4$ numerical orbit integrations in the elliptical-like galactic potential adopted in this work.
}
\label{fig:diag_ellip}
\end{figure*}

\begin{figure*}
   \centering
   \includegraphics[width=\hsize]{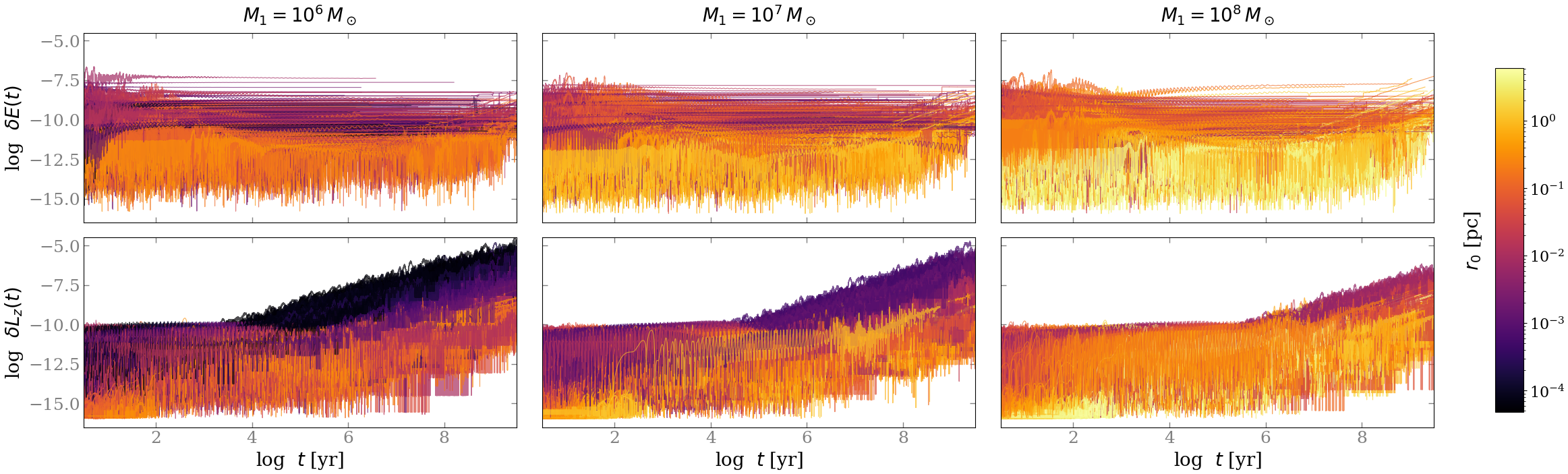}
\caption{
Same as Figure~\ref{fig:diag_ellip}, but for the spiral-like galactic potential.
In this case, the angular-momentum diagnostic corresponds to the fractional variation of the conserved $z$-component, of the angular momentum.
}
\label{fig:diag_sp}
\end{figure*}

Figure~\ref{fig:diag_ellip} shows the fractional variations of energy and angular momentum of the $10^4$ simulated orbits in the elliptical-like host potential leading to Figure~\ref{fig:d_t_M1s}. Each column corresponds to a different value of the central SMBH mass $M_1$, as indicated. Similarly, Figure~\ref{fig:diag_sp} shows the corresponding accuracy diagnostics for the spiral-like host potential, using the fractional variation of $L_z$ instead of the total angular-momentum magnitude.

In both host models, the fractional variations of the conserved quantities remain typically below the $10^{-6}$ level over the full integration time. This demonstrates that the numerical drift in the integrations is negligible compared with the dynamical scales relevant for the post-slingshot trajectories. We therefore conclude that the computed merger offsets are not limited by numerical conservation errors in the orbit integration.


\section*{Declaration of generative AI and AI-assisted technologies in the manuscript preparation process}

\section*{Acknowledgements}
JCRR acknowledges support from brazilian scientific funding agencies FAPERJ (grant E-26/205.635/2022) and FACC (CBPF\_TEMATICO\_2024 - IA, 01.25.0215.00 - Ref. 1021/24). DCR thanks CNPq-Brazil and FAPES-Brazil for partial support.






\bibliographystyle{elsarticle-num-6authors}

\bibliography{refs}






\end{document}